\documentclass[a4paper,11pt]{article}
\pdfoutput=1

\usepackage{jcappub}

\usepackage{amsmath}
\usepackage{amsfonts}
\usepackage{amssymb}
\usepackage{subcaption}
\usepackage{graphicx}
\usepackage{booktabs}
\usepackage{wrapfig}
\usepackage{bm}

\newcommand{\link}[1]{\href{http://arxiv.org/abs/#1}{{\tt #1}}}

\newcommand\be{\begin{equation}}
\newcommand\ee{\end{equation}}
\newcommand{\mrm}[1]{\mathrm{#1}}
\newcommand\cov{\mathbf{C}}
\newcommand\Sig{\mathbf{\Sigma}}
\newcommand\T{\mathrm{T}}
\newcommand{\calL}{\mathcal{L}}
\newcommand{\calN}{\mathcal{N}}
\newcommand{\mat}[1]{{\mathbf{#1}}}

\newcommand\HO{H_0}
\newcommand\LCDM{$\Lambda$CDM}
\newcommand\IXCDM{$q_1$XCDM}
\newcommand\logXCDM{$q$XCDM}
\newcommand{\zdrag}{z_\mathrm{d}}
\newcommand{\Omk}{\Omega_K}
\newcommand{\Oml}{\Omega_X}
\newcommand{\Omb}{\Omega_b}
\newcommand{\Omc}{\Omega_c}
\newcommand{\Omm}{\Omega_m}

\newcommand{\omb}{\omega_b}
\newcommand{\omc}{\omega_c}

\newcommand{\Ombo}{\Omega_{b0}}

\newcommand{\Ommo}{\Omega_{m0}}
\newcommand{\Omlo}{\Omega_{X0}}
\newcommand{\Omko}{\Omega_{K0}}

\newcommand{\qfid}{\bm{q}^\mathrm{fid}}

\title{Probing the independence within the dark sector in the fluid approximation}
\author{Lawrence Dam$,^{1}$ Krzysztof Bolejko$,^{2}$ and Geraint F.~Lewis$^{1}$}
\affiliation{$^{1}$Sydney Institute for Astronomy, The School of Physics, A28, The University of Sydney, NSW 2006, Australia}
\affiliation{$^{2}$School of Natural Sciences, College of Sciences and Engineering, University of Tasmania, Private Bag 37, Hobart TAS 7001, Australia}
\emailAdd{ldam4036@uni.sydney.edu.au}
\emailAdd{krzysztof.bolejko@utas.edu.au}
\emailAdd{geraint.lewis@sydney.edu.au}

\abstract{The standard model of cosmology is based on two unknown dark components that
are uncoupled from each other.
In this paper we investigate whether there is evidence for an interaction
between these components of cold dark matter (CDM) and dark energy (DE).
In particular, we focus on a minimal extension and reconstruct the
interaction history at low-redshifts non-parametrically using a variation
of the commonly used principal component analysis.
Although we focus on the interaction in the dark sector, any significant
deviation from the standard model that changes the expansion history of
the Universe, should leave imprints detectable by our analysis. Thus,
detecting signatures of interaction could also be indicative of other
non-standard phenomena even if they are not the results of the interaction.
It is thus interesting to note that the results presented in this paper do
not provide support for the interaction in the dark sector, although the
uncertainty is still quite large. 
In so far as interaction is present but undetectable using current data,
we show from a Fisher forecast that forthcoming LSST and DESI surveys
will be able to constrain a DM-DE coupling at $20\%$
precision --- enough to falsify the non-interacting scenario, assuming 
the presence of a modest amount of interaction.}

\begin{document}
\maketitle
\flushbottom

\section{Introduction}

The $\Lambda$ Cold Dark Matter (\LCDM) model is now firmly established
as the standard paradigm of cosmology, having fitted a wide range
of observations \cite{Aghanim:2018eyx}. It is, nevertheless, a phenomenological
model and thereby provides no explanation for non-baryonic CDM, and the
cosmological constant that appears to drive cosmic acceleration
\cite{Riess:1998cb,Perlmutter:1998np}.
Although these two ingredients make up $95\%$ of the matter-energy content
of the Universe, little is known about their non-gravitational nature.

In the standard model it is assumed that (\emph{i}) CDM is pressureless
and dilutes with the cosmic expansion as $\rho_c\propto a^{-3}$ and (\emph{ii})
DE has negative pressure, and is undiluting with an equation of state
$w=-1$, i.e.\ a cosmological constant. In the development of the concordance
cosmology the need for these two components arose at
different times, from different lines of evidence, both astrophysical and
cosmological. They are both considered independent
of each other with different evolutions, and thus it is interesting to note that the 
coincidence problem --- i.e.\ the observation that DM and DE have comparable
densities only recently when for most of the lifetime of the Universe they
were different --- occurs roughly when the validity of the fluid approximation
might be questioned.

The fluid approximation, i.e.\ the assumption that the cosmic density field behaves
as an ideal fluid, is used at both early and late times of cosmological
evolution. While this is a reasonable assumption in the former regime, when
gravitational structures had yet to form and matter and energy existed as
a thermal bath of particles, it might be questioned whether the fluid
approximation still holds in the late Universe \cite{Wiltshire:2011vy}. 
From the primordial epoch to the epoch preceding the first gravitational
structures the evolution of the Universe could
simply be described by thermal physics. By contrast, the late Universe we observe
today is significantly more complicated, being composed of a complex hierarchy
of nonlinear gravitational structures.
Tracers of the cosmic density are no longer simply parcels of fluid particles
but are instead galaxies that follow the underlying density field in
nontrivial ways \cite{Desjacques:2016bnm}.

Models that involve interacting dark energy have long been studied
as alternatives to $\Lambda$
(see the reviews \cite{Copeland:2006wr,Bolotin:2015,Wang:2016lxa} and references therein).
Early work centred on quintessence models of DE, in which the scalar field is
coupled to either matter \cite{Amendola:1999er,Chimento:2003iea} or only dark matter
\cite{Zimdahl:2001ar,Farrar:2003uw,Amendola:2003eq}.
If the interaction is contained in the dark sector, however, a particularly
appealing feature is that it can provide an explanation of the coincidence problem
\cite{Zimdahl:2001ar,Chimento:2003iea}.
Recent work though has largely shifted to the question of whether DE is time-varying
(i.e.\ with a possible time-dependent DE equation of state)
and models now tested against data are typically based on the fluid picture of DE
\cite{Sandvik:2002jz,
Bento:2004uh,Barrow:2006hia,Guo:2007zk,Valiviita:2008iv,delCampo:2008jx,CalderaCabral:2009ja,
Majerotto:2009np,Valiviita:2009nu,Boehmer:2008av,Wands:2012vg,Wang:2013qy,
Wang:2014xca,Salvatelli:2014zta,Wang:2015wga,
Nunes:2016dlj,Kumar:2017dnp,DiValentino:2017iww,Yang:2018euj,Yang:2019uzo,Pan:2019gop}.
In all of these models, however, the fundamental mechanism giving rise
to the coupling is generic with only the effects of interaction on linear scales
usually studied (though $N$-body simulations have recently begun
investigating nonlinear scales \cite{Zhang:2018glx,Zhang:2018mlj}).
While the detection of interaction may be physical in nature it could
also be argued to signal a breakdown of the fluid approximation.
Even if the fluid approximation breaks down we might still expect a
fluid-like evolution, just one that is different from the usual scalings.

These models can also be motivated by the fact that gravitational probes
are sensitive only to the \emph{total} energy-momentum tensor $T_{\mu\nu}$
\cite{Kunz:2007rk}, with the splitting of $T_{\mu\nu}$
into different constituents typically based on physical considerations and 
the strength of gravity insensitive to any coupling.
At late times it is reasonable to think ordinary matter and radiation
(baryons, neutrinos, photons etc) in the cosmic fluid do not couple.
Moreover, interactions between Standard Model
particles and the dark sector are strongly constrained by experimental data.
However, without a fundamental theory behind the dark sector there is
{\it a priori} no reason to split into non-interacting DM and DE.

Recently, such models have seen renewed interest
as a possible solution to tensions in the measured values of 
$H_0$ and $\sigma_8$
\cite{Salvatelli:2014zta,Kumar:2017dnp,DiValentino:2017iww,Yang:2018euj,Vattis:2019efj,Pan:2019gop}.
Deviations from the \LCDM\ scenario hint at a breakdown
of the usual assumptions of the dark sector, which could be interpreted in
several different ways. In particular, if we take DM and DE to be as yet undetected 
particles or fields then detection of interaction is to be understood
at face value, i.e.\ physical in nature. Alternatively, if we take the dark
sector to be phenomenological artefacts required for concordance with
observations, then it might call into question the reality of the
dark components.
It is also timely to revisit some of the assumptions of the \LCDM\ model
given that observational cosmology is poised to see an influx
of data from next generation experiments.

Without guidance from theoretical arguments, all parametrisations of the
interaction investigated so far are necessarily phenomenological.
Naturally, simple forms of the interaction have largely been pursued
\cite{Bolotin:2015,Barrow:2006hia,delCampo:2008jx,Nunes:2016dlj,DiValentino:2017iww,Yang:2019uzo}.
Alternatively, we can take a less rigid approach by directly reconstructing
the quantity of interest from, e.g.\ a set of basis functions or Pad\'{e}
approximants. However, this approach of introducing a large number of
degrees of freedom generally suffers from parameter degeneracies leading to
slow convergence using standard parameter estimation methods or weak constraints.
In this work we will take a more data-driven approach. We
use a generalised principal component analysis, allowing us to reconstruct the
interaction history from a statistically decorrelated basis of eigenfunctions. The
main advantage is that we are able to constrain only those features that are
actually being probed, and thus let the data decide on the best functional form
for the interaction.

The plan of this paper is as follows. In Section \ref{sec:prelim} we 
review aspects of interacting models, discuss the relevant theory, and set out the
model to be analysed. In Section \ref{sec:method} we describe the
data, statistical methods and tools used in the analysis. Section \ref{sec:analysis}
presents the results; in Section \ref{sec:forecast} we study the detectability of
interaction in upcoming surveys; finally, in Section \ref{sec:concl} we 
summarise our main findings.

\section{Preliminaries \label{sec:prelim}}
We consider an energy-momentum tensor consisting of multiple fluids,
labelled $A$. 
Typically it is assumed that each fluid species $A$ satisfies its own
energy-momentum conservation equation, $\nabla_\mu T^{\mu\nu}_{(A)}=0$. 
In general, if we allow the transfer of energy-momentum between species then
\be\label{eq:Tmunu}
\nabla_\mu T^{\mu\nu}_{(A)} = Q^\nu_{(A)},
\ee
where $Q^\nu_{(A)}$ is the covariant interaction of species $A$. The
conservation of the total energy-momentum $T^{\mu\nu}=\sum_A T^{\mu\nu}_{(A)}$ 
implies the balance condition $\sum_A Q^\nu_{(A)} = 0$.
Separating the energy and momentum part by decomposing $Q^\nu_{(A)}$
relative to the fluid 4-velocity $u^\mu_{(A)}$ of each fluid component we 
can write
\be\nonumber
Q_{(A)}^\nu= Q_{(A)} u_{(A)}^\nu + f_{(A)}^\nu,
	\qquad g_{\mu\nu}f_{(A)}^\nu u_{(A)}^\nu=0,
\ee
where $Q_{(A)}$ is the rate of energy transfer, $f^\nu_{(A)}$ the
rate of momentum transfer, $u^\nu_{(A)}$ the 4-velocity,
and $g_{\mu\nu}$ is the metric tensor. It can be observed that
$Q^\mu$ does not appear in Einstein's equations as they depend on
$T_{\mu\nu}$ and not its derivative. The interaction term enters
through the fluid equations only, which are in general modified from there usual
forms.

As we consider only interaction between CDM ($c$) and
DE ($X$) we have $Q^\nu:=Q_{c}^\nu=-Q_{X}^\nu$. Following
\cite{Valiviita:2008iv} the form of the covariant interaction we assume
to be
\be\label{eq:Q-geo}
Q^\mu=Qu^\mu_{c},
\ee
where $Q$ is time-dependent only and $u^\mu_{c}$
is the 4-velocity of CDM. In this simple model there is no net
momentum transfer in the rest frame of DM; any transfer that takes
place is along the geodesic flow of DM. Consequently, in the
synchronous gauge the peculiar velocity of DM (and also baryons)
vanishes and we have $u_c^\mu=(1,0,0,0)$.
Moreover, as there is no momentum
transfer, no spatial gradients arise in the density of DE and we 
have $\delta\rho_X=0$, i.e.\ DE is spatially homogeneous. 
This model is known as the \emph{geodesic interaction model} and is among
the simplest interacting extension to \LCDM.
In this model the fluid equations retain their usual non-interacting forms
and it is the interaction model we consider in this work.

\subsection{Model specifications\label{sec:model}}
The form of the function $Q$ in \eqref{eq:Q-geo} we assume to be of the
following form
\be\label{eq:Q}
Q(a)=q(a)H(a)\rho_{X}(a),
\ee
where $q(a)$ is the dimensionless interaction history, $H(a)$ is the Hubble
constant and $\rho_{X}(a)$ is the DE density. The chosen form
for $Q(a)$ is for convenience only; any arbitrariness of $Q(a)$ is absorbed
into $q(a)$. Note however that it does not necessarily lead to solutions
that remain physical into the future. For instance, if $q>0$ then DM 
decays unbounded at a rate proportional to
the DE density, eventually going negative. Therefore, we consider
\eqref{eq:Q} an ansatz valid for the late-epoch that we focus on in this work.

Given that the matter density and DE density are
approximately equal in the recent past we expect that any interaction
will be greatest at low-redshifts. This is generally realised by assuming
a logistic-like interaction parametrisation in which $Q$ only
becomes appreciable at late-times when $\rho_X\simeq \rho_c$.

The convention we use here is that positive values of $Q$ (or $q$) gives
a universe in which DM decays to DE, while negative values gives the reverse
behaviour. For a fixed $\Ommo$, models with $q>0$ will have a greater
fraction of matter through all epochs.
In such a case, the growth of structure is enhanced relative to
\LCDM, as the universe is more matter dominated than in the
non-interacting case. For $q<0$ the growth of structure is suppressed
relative to \LCDM\ as the onset of DE occurs earlier.

Assuming a DE equation of state $w_X=-1$, the coupled
continuity equations becomes
\be\label{eq:dot-rho-int}
\dot\rho_c+3H\rho_c=-Q,
\qquad \dot\rho_X = Q,
\ee
where overdots denote differentiation with respect to cosmic time.
The cosmological constant is of course recovered when $Q=0$ so
$\rho_{X}\propto\Lambda$. All other fluid components evolve in the
standard, non-interacting way. 

The issue of how to parametrise $q$ can be likened to that of
determining the DE equation of state. 
In the absence of any plausible $q$ from theory we will take a
model-independent approach and reconstruct it directly from data.
We divide $q$ up into $n$ bins and constrain the amplitudes $q_i$ of
each bin.
The bins are chosen to be uniformly spaced in scale factor $a$,
with edges $a_0<a_1<a_2<\ldots<a_n$.
The $i^\mrm{th}$ bin spans the interval $[a_{i-1},a_i)$ and we set
$a_n=a_\mrm{max}=1$. Since $q(a)$ has a piecewise constant amplitude in
each bin we represent it as
\be\label{eq:q}
q(a)=\sum_{i=1}^n q_i T_i(a),
\qquad 
T_i(a)=\begin{cases}1,&a_{i-1}\leq a<a_i,\\0,&\text{otherwise}.\end{cases}
\ee
The amplitudes $q_1,q_2,\ldots,q_n$ are dimensionless parameters
characterising the interaction strength. We set $q(a)=0$ outside the
binning range $[a_0,a_n]$. The low-redshift window chosen is motivated
by the fact that the onset of cosmic acceleration occurs in the recent
past.

The Friedmann equation is given by
\be
H^2(a)/\HO^2=\Omm(a)+\Oml(a) + \Omk(a),
\ee
where $H_0$ is the present-day value of the Hubble constant,
$\Omk(a)=\Omko a^{-2}$ with $\Omko$ the spatial curvature parameter, and
$\Omm(a)=\Omb(a)+\Omc(a)$ together with $\Oml(a)$
have modified time-dependence given by solving \eqref{eq:dot-rho-int}.
With the specific form given by \eqref{eq:Q} we find
\begin{align*}
\Oml(a)&=\Omlo \left(\frac{a}{a_{j-1}}\right)^{q_j} 
		\prod_{i=1}^{j-1} \left(\frac{a_i}{a_{i-1}}\right)^{q_i},\\
\Omm(a)
&=\Ommo a^{-3} + \Omlo\sum_{i=1}^{j} \frac{q_i}{q_i+3} 
	\left[\prod_{k=1}^{i-1} \left(\frac{a_k}{a_{k-1}}\right)^{q_k}\right]
\times
\begin{cases}
    \bigg(\dfrac{a_{i-1}}{a}\bigg)^3
 - \bigg(\dfrac{a_i}{a_{i-1}}\bigg)^{q_i} \bigg(\dfrac{a_i}{a}\bigg)^3, & i<j, \\[15pt]
    \bigg(\dfrac{a_{i-1}}{a}\bigg)^3
            - \bigg(\dfrac{a}{a_{i-1}}\bigg)^{q_i}, & i=j,
\end{cases}
\end{align*}
where $a\in [a_{j-1},a_j)$, $\Ommo$ is the present matter density parameter,
and $\Omlo$ is the present DE density parameter.
If $a<a_\mrm{min}=a_0$ then the densities recover their usual forms,
$\rho_c\propto a^{-3}$ and $\rho_X=\mrm{const}$.
The binning strategy is chosen to effectively impose standard \LCDM\
evolution, up until the onset of cosmic acceleration at late-times where
we are most interested.

At the level of perturbations the presence of interaction modifies
the continuity equation to allow an exchange of energy between fluid
species. For the total matter fluctuation $\delta_m$ we have,
in the synchronous gauge,
\be\label{eq:dot-deltam}
\dot\delta_m +\frac{1}{2}\dot{h} = (Q/\rho_m)\delta_m,
\ee
where $\rho_m=\rho_b+\rho_c$ is the total matter density in the background.
From Einstein's equations, the metric perturbation $h$ satisfies
\be\label{eq:h-ode}
\ddot{h} + 2H\dot{h} = -8\pi G (\delta\rho+3\delta p),
\ee
where $G$ is the gravitational constant, $\delta\rho=\sum_A\delta\rho_{(A)}$
and $\delta p=\sum_A\delta p_{(A)}$ are the total density and pressure
fluctuations, respectively.
An ordinary, second-order differential equation can be obtained from
\eqref{eq:dot-deltam} and \eqref{eq:h-ode} that is closed in the
perturbation variable $\delta_m$:
\be\nonumber
\ddot\delta_m
+\left(2H-Q/\rho_m\right)\dot\delta_m
-\left[2HQ/\rho_m+\frac{d}{dt}(Q/\rho_m)\right]\delta_m
=4\pi G(\delta\rho+3\delta p).
\ee
At late-times the effect of radiation is negligible so we neglect it
and treat the total density perturbations as composed of baryons, CDM and DE.
In comoving synchronous gauge it can be shown that $\delta\rho_{X}=0$,
i.e.\ dark energy is spatially homogeneous and nonclustering \cite{Wands:2012vg}.
Equation \eqref{eq:dot-deltam} implies the amplitudes of fluctuations evolve
uniformly. Thus we can write $\delta_m(\bm{x},a)=D(a)\delta_m(\bm{x},a=1)$,
where $D(a)$ is the growing mode.
With a change of independent variable $t\to a$ the above equation becomes
\be\label{eq:growth-eq}
D''
	+\frac{1}{a}\left(3+\frac{d\ln H}{d\ln a}-\Gamma\right)D'
	=\frac{1}{a^2}\left[\frac{3}{2}\Omm(a)
	+\frac{1}{H}\frac{d}{d\ln a}\big(\Gamma H\big)+2\Gamma\right] D,
\ee
where primes denotes differentiation with respect to scale factor,
and we have defined the dimensionless function $\Gamma\equiv Q/(\rho_m H)$
and $\Omm(a)\equiv 8\pi G\rho_m(a)/3H^2(a)$.
Note we have suppressed all dependence on the scale factor, except for
$\Omm(a)$, to avoid confusion with the matter density parameter $\Ommo$.
We solve this equation numerically with the initial conditions
$D(a_\mrm{init})=a_\mrm{init}$ and $D'(a_\mrm{init})=1$, taking
$a_\mrm{init}=0.03$ (or $z\simeq30$), that is we begin integration at a time
deep in the matter-dominated epoch when
$D\sim a$. Having then solved \eqref{eq:growth-eq} we normalise the growth
factor to unity today. The quantity of interest is not $D(a)$, which is
not observable, but the growth rate $f\equiv d\ln D/d\ln a$. Without
solving this equation we can understand the effect of interaction qualitatively.
If $Q>0$ there are two competing effects:
on the one hand a conversion of DM to DE produces a faster cosmic
expansion and a lower matter fraction but on the other we have
$\Gamma>0$ driving the dissipation down and reinforcing the source term.
In effect this means we can always compensate a high $\Ommo$ with a low $q$.
Clearly, $D(a)$ no longer just depends on the cosmic expansion $H(a)$, but 
also $Q(a)$. This implies that a detection of non-zero $Q$ can be translated to
mean a violation of a consistency relation if the assumptions of \LCDM\
do not hold \cite{Knox:2005rg,Ishak:2005zs,Chiba:2007rb}.

\section{Methodology\label{sec:method}}
\subsection{Data sets}
In this section we describe the data used and emphasize the
physics they probe.

\paragraph*{Baryon Acoustic Oscillations.}
Galaxy surveys exhibit enhancements at a certain length scale
in the clustering of matter, due to Baryon Acoustic Oscillations (BAO).
From the anisotropic 2-point correlation function BAO surveys
commonly report the distilled quantity
\be
d_z(z)\equiv r_s(\zdrag)/D_V(z),\qquad
D_V(z)\equiv \big[(1+z)^2 d_A^2(z) \times cz/H(z) \big]^{1/3},
\ee
where $z$ is an effective redshift found by taking a weighted average
of galaxies in a given slice, $\zdrag$ is the redshift at the baryon
drag epoch, $d_A(z)=(1+z)^{-1}\int^z_0dz'/H(z')$
is the angular diameter distance for a spatially flat
FLRW metric, and for a sound speed $c_s(z)$ in the baryon-photon fluid 
$r_s(z)=\int^\infty_z dz'\,c_s(z')/H(z')$ is the comoving sound horizon.
For the particular interacting model being considered the sound speed
remains unchanged from \LCDM\ (see Appendix \ref{FF}).

We use data from the Six-degree Field Galaxy Survey (6dFGS) \cite{Beutler:2011hx}, 
the Sloan Digital Sky Survey (SDSS) DR7 Main Galaxy Sample \cite{Ross:2014qpa},
the LOWZ and CMASS galaxy samples of
the Baryon Oscillation Spectroscopic Survey (BOSS) DR12 \cite{Cuesta:2015mqa},
and the SDSS Luminous Red Galaxies (LRG) \cite{Padmanabhan:2012hf}.
We also use the three correlated measurements reported by WiggleZ
with the provided covariances \cite{Blake:2011en}. The data comprise
eight measurements at different effective redshifts of the distilled
parameter given variously as $d_z$, its reciprocal $D_V(z)/r_s(\zdrag)$,
or sometimes normalised to a fiducial cosmology.

\paragraph*{Redshift Space Distortions.}
The growth of structure in the Universe depends on its energy contents
through its effect on background expansion (and importantly also interaction).
The competition between 
cosmic expansion and the tendency for inhomogeneous regions to be further
enhanced by gravity gives a useful dynamical probe at linear scales,
particularly of DE and modified gravity.
Galaxy surveys exploit the anisotropies induced on the power spectrum
from redshift space distortions (RSD) to measure the normalised
growth rate $f\sigma_8$, where $\sigma_8(z)$ is the root-mean-square
of the amplitude of matter fluctuation averaged in a spherical volume
of radius $8h^{-1}\,\textrm{Mpc}$.
In the linear regime $\sigma_8(z)$ simply scales with $D(a)$
so $\sigma_8(z)=D(a)\sigma_{80}$, where $\sigma_{80}$ is its present-day value
and can be treated as a free parameter.

In this analysis we use the same data as compiled by Planck (2018)
\cite{Aghanim:2018eyx}, which consists of measurements
from 6dFGS \cite{Huterer:2016uyq,Beutler:2012px}, SDSS MGS \cite{Howlett:2014opa},
SDSS DR7 LRG \cite{Oka:2013cba}, GAMA \cite{Blake:2013nif},
BOSS DR12 \cite{Alam:2016hwk}, WiggleZ \cite{Blake:2012pj},
VIPERS \cite{Pezzotta:2016gbo}, FastSound \cite{Okumura:2015lvp},
and BOSS DR14 quasars \cite{Zarrouk:2018vwy}.

\paragraph*{Type Ia Supernovae.}\label{type-ia-supernova}
Type Ia supernovae (SNe Ia) are standardisable candles that can be used
to probe the expansion history. The distance to SNe Ia is
given by the distance modulus defined as
\be\label{eq:distmod}
\mu(z) = m - M = 5\log_{10}\left[d_L(z)/\text{10 pc}\right],
\ee
where $m$ is the apparent magnitude, $M$ is the absolute
magnitude, and $d_L(z)=(1+z)^2 d_A(z)$ is the luminosity distance.

The cosmological observable is the distance moduli $\mu(z)$, which is
to be compared to the measured value given by the Tripp relation
\be\label{eq:tripp}
\mu = m_B^* - M + \alpha x_{1} -\beta c,
\ee
with the peak apparent magnitude $m_B^*$ and intrinsic magnitude $M$ are
given in $B$-band. The two additional terms --- known as Phillips corrections --- 
are the time stretch parameter ${x}_1$ and the colour correction parameter
at maximum brightness ${c}$ \cite{Phillips:1993ng,Phillips:1999vh}.
The stretch and colour coefficients, $\alpha$ and $\beta$, are nuisance 
parameters, i.e.\ parameters to be fitted simultaneously with the cosmological
parameters.

The data for each SN Ia are the parameters $\hat{m}^*_B$, $\hat{x}_1$
and $\hat{c}$ as well as the heliocentric redshift $\hat{z}$ produced using
the SALT2 light-curve fitting procedure \cite{Guy:2007dv}. 
We use the Joint Light-curve Analysis (JLA) catalogue of $740$
spectroscopically confirmed SNe Ia with redshifts ranging from $z=0.01$
to 1.3 \cite{Betoule:2014frx}.%
\footnote{\url{http://supernovae.in2p3.fr/sdss_snls_jla/}}
Though the more recent Pantheon sample \cite{Scolnic:2017caz} is larger,
with over 1000 SNe Ia, it does not yet include ancillary data products
we need for the statistical method used here (described in Section \ref{sec:SM}).

\paragraph*{Cosmic Chronometers.}
The basic idea of the cosmic chronometers (CC) as a probe is to directly
measure the cosmic expansion history by using the kinematic form
$H(z)=-(1+z)^{-1}dz/dt$ (assuming a Friedmann-Lema\^{i}tre-Robertson-Walker
(FLRW) background redshift $1+z=1/a$). In principle, because it is not an
integrated quantity like $d_L(z)$ it provides greater sensitivity to cosmological parameters.
The difficulty however is in obtaining accurate estimates of the differential changes
in redshift. At present, measurements are at about $6\%$ precision and limited
by systematics. Observations of the age differences of passively evolving old,
elliptical galaxies are typically serve as the ``standard clocks'' of this method.
This analysis makes use of 26 measurements of $H(z)$ with redshifts ranging
from $z\simeq 0.01$ to $z\simeq 2$
\cite{Moresco:2012jh,Moresco:2012by,Moresco:2015cya,Moresco:2016mzx}.

\paragraph*{Cosmic Microwave Background.}
To complement the low-redshift probes we also include data from Planck.
Although the CMB is mainly a probe of the early Universe (when dark
energy was negligible), it does contain some valuable geometric information
through the angular scale of the sound horizon \cite{Vonlanthen:2010cd}.
At the background level, the CMB data provides a precise determination
of the distance to last scattering.

We include Planck data using the compressed CMB likelihood
method that considers a handful of parameters (thought of as observables)
that summarise key features of the CMB power spectrum
\cite{Kosowsky:2002zt,Wang:2007mza,Mukherjee:2008kd}. 
The data consists of the following:
(\emph{i}) the CMB shift parameter
$\mathcal{R}\equiv \sqrt{\Omm \HO^2}D_A(z_*)$,
where $D_A(z)\equiv(1+z)d_A(z)$ is the comoving angular
diameter distance evaluated at the redshift of last scattering $z_*$;
(\emph{ii}) the angular scale of the sound horizon at last scattering,
$\ell_A \equiv \pi D_A(z_*)/r_s(z_*) = \pi/\theta_*$,
where $r_s(z_*)$ is the comoving sound horizon
and $\theta_*=r_s(z_*)/D_A(z_*)$ is the angular size of the sound
horizon; (\emph{iii}) the physical baryon density $\omb$.
With a precision of $\lesssim1\%$, these three quantities
are among the most precisely determined by Planck.
As they summarise key geometric features of the CMB angular
power spectrum they are sometimes referred to as the CMB distance prior.
The observable $\mathcal{R}$ determines the distance to the last scattering
surface independent of $\HO$, $\ell_A$ is closely
related to the position of the first acoustic peak, and $\omb$
sets the relative heights of odd to even peaks.
When combined with other data sets there is no significant loss of 
information using the compressed likelihood versus the full
likelihood \cite{Mukherjee:2008kd}.
Although they are not strictly cosmology-free measurements, but
rather constrained quantities obtained from a CMB analysis assuming
a given model, they can be considered early Universe observables
independent of the (late-)DE model assumed \cite{Ade:2015rim}.
We use the Planck 2015 data release \cite{Ade:2015rim} of the lensing
amplitude marginalised-compressed likelihood.%
\footnote{It is well-known that combinations of $\Omega_\Lambda h^2$
and $\Omko h^2$ that give the same $\mathcal{R}$ will produce a
near identical CMB spectrum at high multipoles \cite{Efstathiou:1998xx}.
However, with constraints from CMB lensing this degeneracy is broken.}
This likelihood is summarised in Table \ref{table:cmb}.

\begin{table}
\begin{center}
\begin{tabular}{r|cc|ccc}
& & & \multicolumn{3}{c}{Correlation coefficients} \\
& $\hat{X}$ & $\sigma$ & $\mathcal{R}$ & $\ell_A$ & $\omb$ \\
\hline
$\mathcal{R}$ & 1.7382& 0.0088  & 1.0 & 0.64 & -0.75 \\
$\ell_A$ & 	  301.63  & 0.15    & - & 1.0 & -0.55 \\
$\omb$ &      0.02262 & 0.00029 & - & - & 1.0 \\
\end{tabular}
\caption{Summary statistics of the compressed CMB likelihood \cite{Ade:2015rim},
		describing the marginalised mean values and their $68\%$ confidence limits.
		The last three columns give the pairwise correlations.}
\label{table:cmb}
\end{center}
\end{table}

\subsection{Statistical modelling\label{sec:SM}}
The parameters are sampled from the posterior formed from the
joint likelihood of BAO, CMB, CC, RSD, and SNe Ia.
For BAO, CC, and RSD we take the likelihoods to be Gaussian distributed
in the data.
As the Planck joint posterior distribution of the data
$\hat{\bm{X}}=(\mathcal{R},\ell_A,\omb)^\T$ is near-Gaussian we
also take the CMB likelihood to be Gaussian
with mean given by the second column of Table \ref{table:cmb} and
covariance matrix
$\Sig_\mrm{CMB}=\Sig^{1/2} \mat{D}\Sig^{1/2}$,
with
$\Sig^{1/2}\equiv\mrm{diag}(\sigma_{\mathcal{R}},\sigma_{\ell_A},\sigma_{\omb})$
and $\mat{D}$ the correlation matrix.

The SN data consists of standardisation outputs from the SALT2 template
and as a result the regression model demands a more principled approach.
We adopt a recent Bayesian hierarchical approach in which the
dependencies are constructed within a probabilistic framework
and has been shown to also deliver tighter constraints on parameters
\cite{Gelman:2014,March:2011xa}.

In the hierarchical approach we introduce the latent variables
$\bm{M}$, $\bm{x}_1$, $\bm{c}$ and $\bm{z}$ (represented as vectors)
as the true variables that we do not observe. 
The SN likelihood involves additional hyperparameters that describe the 
distributions of latent or unobserved variables relating to the colour,
stretch, and absolute magnitude.
These distributions are taken to be Gaussian with means and standard
deviations included in the set of free parameters $\theta$.
The likelihood of the SALT2 outputs has the probabilistic form
$\mathcal{L}_\mrm{SN}({\theta};D)
\equiv p(\hat{\bm{z}},\hat{\bm{m}}_B^*,\hat{\bm{x}}_1,\hat{\bm{c}} \,|\, {\theta})$.
The set of parameters ${\theta}$ includes the nuisance
parameters $\alpha$ and $\beta$.
The data $D$ includes $\hat{\bm{m}}_B^*$, $\hat{\bm{x}}_1$, $\hat{\bm{c}}$
and $\hat{\bm{z}}$.
The details of the marginalised likelihood used in this analysis are
given in Appendix \ref{SN-BHM}.

To summarise, our regression model consists of the
base cosmological parameters $\Ommo$, $\Ombo$, $\sigma_{80}$ the
reconstruction parameters $q_1,\ldots,q_n$,
the SN nuisance parameters $\alpha$, $\beta$ and additional SN parameters 
$M_0,x_*,c_*, R_M^2,R_x^2,R_c^2$ that describe the Gaussian priors
of the latent variables.

\subsection{Priors}
\subsubsection{Smoothing priors on reconstruction}
Given the emphasis of this analysis on the parameters $q_1,q_2,\ldots,q_n$
it is important that appropriate priors are chosen.
Typically flat priors are used to allow the inference to be driven by the
data and provided the likelihood is informative and supported by the prior
this is usually a reasonable choice. However in reconstruction there are necessarily a
large number of degrees of freedom and these flat priors become
informative relative to the likelihood. For sparse data it is unlikely that all
parameters can
be constrained and, because neighbouring bins are uncorrelated, often leads
to a noisy reconstruction. Increasing the number of bins only introduces
more unconstrained degrees of freedom and as a result the posterior is
multimodal and convergence of MCMC methods is slow. This is one of the basic
problems that reconstruction methods face.
One popular method is to add a regularisation or penalty term to the
$\chi^2$, designed in a way to favour smooth reconstructions. The method
we adopt here instead incorporates this information into the prior
\cite{Crittenden:2005wj}. The prior is assumed to be of Gaussian form
\be\label{eq:pi-q}
\pi(\bm{q})
=\frac{1}{(2\pi)^{n/2}\sqrt{\det\cov_\pi}}
	\, \exp\left[-\frac{1}{2}(\bm{q}-\qfid)^\T \cov^{-1}_\pi
				(\bm{q}-\qfid)\right],
\ee
where $\bm{q}=(q_1,q_2,\ldots,q_n)^\T$ and similarly for the fiducial
model $\qfid$. The off-diagonal entries of $\cov_\pi$
can be specified so as to favour smoothness, ensuring that the
$q_i$s do not change abruptly between bins. Since the discretisation of the
unknown $q(a)$ is achieved by a simple average over each bin we can write
\be
q(a)\to q_i=\int^1_0 da \,W_i(a) q(a),
\ee
where $W_i(a) = T_i(a)/(a_i-a_{i-1})$ is the normalised top-hat function.
With this the covariance matrix can be written in component form as
\begin{align}
(\cov_\pi)_{ij}
&\equiv \left\langle(q_i-q^\mrm{fid}_i)(q_j-q^\mrm{fid}_j)\right\rangle \nonumber\\
&= \int^1_0 da\,W_i(a)\int^1_0 da'\,W_j(a')
	\left\langle(q(a)-q^\mrm{fid}(a))(q(a')-q^\mrm{fid}(a'))\right\rangle.
	\label{eq:Cij}
\end{align}
Thus covariances between bins are encoded in temporal correlations of $q(a)$
given by the two-point correlation function:
\be
 \xi(a,a')\equiv\left\langle
  \big(q(a)-q^\text{fid}(a)\big)\big(q(a')-q^\text{fid}(a')\big)\right\rangle.
\ee
Here $\xi(a,a')$ is a function that we are free to specify. As a matter
of convenience we assert that correlations are invariant under time
translations and reversals, which implies that the arguments of $\xi$
depends only on the magnitude of the difference between $a$ and $a'$,
i.e.\ $\xi(a,a') = \xi(|a-a'|)$.
A physically plausible $\xi$ should take into
account the fact that correlations should be strongest
for small separations then fall off with ``distance''.
In this analysis we use the CPZ correlation function \cite{Crittenden:2005wj},
which effects  a $\sim1/r^2$ fall-off:
\be\label{eq:xi}
\xi(|a-a'|) = \frac{\xi_0}{1+(|a-a'|/a_c)^2}.
\ee
The tuning parameter $a_c$ sets the characteristic correlation length
while $\xi_0$ determines the overall strength of correlations. 
Larger values of $a_c$ correspond to stronger correlations between
bins, vice-versa, and in the limit $a_c\to0$ there are no cross-correlations
between bins, which implies ${\xi(|a-a'|)\to\delta_D(a-a')}\xi_0$. On the other
hand, in the limit $a_c\to\infty$ the denominator approaches unity and
${\xi(|a-a'|)\to\xi_0}$ for all $a$ and $a'$, and we effectively recover
a flat prior.

In this analysis we consider two fiducial models: (\emph{i}) $\qfid$
determined by a five bin running average of $\bm{q}$, which we will call
Prior I and 
(\emph{ii}) a \LCDM-biased prior with $\qfid=0$, which we call Prior II.
We remark that in using Prior II we can write $\qfid=\mat{R}\bm{q}$,
for some constant matrix $\mat{R}$, i.e.\ $\qfid$ is now a function of $\bm{q}$.
Technically, the PDF \eqref{eq:pi-q} is not normalised to
unity upon inserting $\qfid=\mat{R}\bm{q}$, but instead must be rescaled by
multiplying it by $\det[(\mat{I}-\mat{R})(\mat{I}-\mat{R})^\T]$.
The prior remains Gaussian but is now centred about $\bm{q}=0$ with a
new covariance matrix that more easily allows the recovery of
low-frequency features.

\subsubsection{Other priors}
For the rest of the parameters, the prior distributions and ranges used in
this analysis are as follows:
the matter density parameter $\Ommo\in[0,1]$, the DE density
parameter $\Omlo\in[0,1]$, the baryon density parameter $\Ombo\in[0,0.4]$,
the log of the clustering amplitude $\log_{10}\sigma_{80}\in[-5,2]$ (a log-uniform
prior of $\sigma_{80}$). For the SN-specific parameters we choose the
intervals $\alpha\in[0,1]$, $\beta\in[0,4]$, $\log_{10} R_M^2\in[-10,4]$,
$\log_{10} R_x^2\in[-10,4]$, $\log_{10} R_c^2\in[-10,4]$; for the means
we choose $M_0\sim\calN(19.3,2^2)$, $x_*\sim\calN(0,10^2)$ and
$c_*\sim\calN(0,1^2)$.

\subsection{Identifying a data-oriented basis}
The binning of $q(a)$ introduces a large number of extra parameters and
it is often the case that the data can support only a few parameters, with
the rest being unconstrained. In such cases a common approach to reconstruction
is to perform a principal component analysis (PCA) to tease out the
features most sensitive to data (usually at low redshifts where data
are more abundant). 
The PCA approach has been widely used to reconstruct
the DE equation of state using real data and mock data in forecasting
exercises
\cite{Huterer:2002hy,Huterer:2004ch,Crittenden:2005wj,Sarkar:2007tx,
Clarkson:2010bm,Crittenden:2011aa,Zhao:2012aw,Said:2013jxa,Zhao:2017cud}.
The principal components (PCs) represent orthogonal directions in parameter space
and therefore provide a statistically decorrelated basis. These directions
are obtained by performing an eigendecomposition of the Fisher matrix, either
of the prior distribution for constraint forecasting or the posterior when
using real data.
When considered as an expansion in the top-hat basis the PCs are
eigenfunctions constructed from an average of $q(a)$ weighted according
to those features that are actually being probed by the data.
In this case the number of bins $n$ is usually
taken to be large so as to allow the recovery of smooth PCs. Using
the correlated prior \eqref{eq:p-q-int} allows
us to employ a greater number of bins than would otherwise be possible,
and effectively smooths out the discrete artefacts that tend to arise
when using real data. In this approach
the $n$ top-hat basis functions are considered as an intermediate basis
from which we construct an uncorrelated basis.

Typically a basis is obtained by finding eigenvectors of the posterior
parameter covariance $\cov_p$. When using the correlated prior \eqref{eq:pi-q}
this approach suffers from an inherent difficulty in finding a natural
way to order the modes, as they do not cleanly separate according to their eigenvalues.
The modes with the largest eigenvalues generally are unconstrained high-frequency modes,
and the number of modes that are actually probed by the data are sensitive
to the chosen $\xi_0$. 

In order to identify a new basis that captures the features the data are 
probing we find it useful to expand $q(a)$ in terms of a basis
expressing the signal-to-noise. This is found by solving the
Karhunen-Lo\`{e}ve (KL) eigenvalue problem \cite{Tegmark:1996bz}
\be\label{eq:geig-eq}
\mat{F}_\pi \, \mathbf{v}_i = \lambda_i \, \mat{F}_p \, \mathbf{v}_i,
\ee
with the Fisher 
information matrices $\mat{F}_\pi=\cov^{-1}_\pi$ and $\mat{F}_p=\cov^{-1}_p$
of the of the prior ($\pi$) and posterior ($p$), respectively. ($\cov_\pi$
can be computed directly from \eqref{eq:Cij} but $\cov_p$ is estimated
from the MCMC samples.)
The set of generalised eigenvectors $\{\bm{v}_i\}$ form the new basis and
we order them from highest to lowest signal-to-noise, which we define here as
$(S/N)_i =\lambda_i^{-1}$, for noise given by $\mat{F}_\pi$ and signal by $\mat{F}_p$.
Because low-frequency modes have the highest $S/N$, we can consider
\eqref{eq:geig-eq} a low-pass filter.
In the space of functions spanned by $T_i(a)$ we can write \eqref{eq:q} in
this basis as
\be\label{eq:q-KL}
q(a)=\sum_{i=1}^n \alpha_i e_i(a),\qquad e_i(a)=\sum_{j=1}^n A_{ij} \, T_j(a),
\ee
where $A_{ij}=(\mathbf{v}_{i})_j$, i.e.\ the rows of $\mat{A}=(A_{ij})$ are
given by the generalised eigenvectors $\mathbf{v}_i$. Note $\{\mathbf{v}_i\}$ form a complete spanning
set of the original basis but are not mutually orthogonal. The basis coefficients
$\alpha_i$ however are uncorrelated by virtue that its covariance matrix is
$(\mat{A}\mat{F}_p\mat{A}^\T)^{-1}=\mat{I}$, where $\mat{I}$ is the identity
matrix. It can also be seen that all $\alpha_i$
have unit variance and their values can be recovered from $\bm{\alpha} = \mat{A}^{-\T}\bm{q}$.

It is important to quantify how many $\alpha_i$ are really being constrained
by the data to avoid fitting for the noise of the reconstruction. To do this
we compute the Bayesian complexity \cite{Spiegelhalter:2002}
\be
C \equiv -2\big(D_\mrm{KL}[p,\pi]-\widehat{D}_\mrm{KL}\big),
\ee
where, for a given posterior $p(\theta|D)$ and prior $\pi(\theta)$,
\be\label{eq:DKL}
D_\mrm{KL}[p,\pi] = \int d\theta \, p(\theta|D) \ln\left(\frac{p(\theta|D)}{\pi(\theta)}\right),
\ee
is the Kullback-Leibler divergence, which quantifies the amount of
information gained from the data, and $\widehat{D}_\mrm{KL}$ is a point
estimator of $D_\mrm{KL}[p,\pi]$. 
The Bayesian complexity effectively tells us how many parameters are being
constrained by the data. Restricting to the parameters
$\theta=\{q_1,q_2,\ldots,q_n\}$ by marginalising over all others, since 
$\pi(\theta)$ is a Gaussian we can write the effective number of $\alpha_i$
being constrained as
\be\label{eq:BC}
C = n-\mrm{tr}(\mat{F}^{-1}_p \, \mat{F}_\pi) 
= n - \sum_{i=1}^n \lambda_i
= n - \sum_{i=1}^n \frac{1}{(S/N)_i}.
\ee
If the $S/N$ is high for all modes then the sum is approximately zero and
the effective number of parameters is equal to the total number of model
parameters.
It should be noted however that the formula defined by the first equality 
of \eqref{eq:BC} only holds if both 
the prior and posterior distributions are Gaussians \cite{Kunz:2006mc}.
Note that in the case of Prior I, the prior Fisher matrix $\mat{F}_\pi$
is not the same as the inverse of \eqref{eq:Cij} but is instead given by
$\mat{F}_\pi=(\mat{I}-\mat{R})^\T \cov_\pi^{-1}(\mat{I}-\mat{R})$.

\section{Analysis\label{sec:analysis}}

In this section we analyse an interacting
model with 20 uniformly spaced bins between $a_\mrm{min}=0.4$ and
$a_\mrm{max}=1$. The reconstruction
of the interaction is thus restricted to this range, which corresponds to
non-uniformly spaced bins from a redshift of $z_\mrm{min}=0$ to $1.5$.
Bins containing no data provide little information on the corresponding $q_i$
so to ensure that each bin contains at least one data point we have
set $a_\mrm{min}=0.4$.
For the correlated $\bm{q}$ prior we choose a smoothing length of $a_c=0.12$,
corresponding to a characteristic correlation with the nearest $\sim 4$ bins.
We will consider $\xi_0=0.2$, which corresponds
to a standard deviation of $q(a)$ when averaged between $a=0.4$ and $a=1$ of
$\sigma_{\bar{q}}\approx 0.4$.

The parameters are estimated from the joint posterior given by 
Bayes's theorem $p({\theta}|D) \propto \calL({\theta};D)\pi({\theta})$,
with $\pi({\theta})$ being the joint prior and $\calL({\theta};D)$
the joint likelihood formed from the BAO, CMB, CC, RSD, and SN
likelihoods. For all data except SN (see Appendix \ref{SN-BHM}) we take
the data to be Gaussian distributed. The parameter set consists
of the following parameters:
\be
\theta=\{\Ommo,\Ombo,\sigma_{80},q_1,q_2,\ldots,q_{20},\alpha,\beta,x_*,c_*,M_0,R_x,R_c,R_M\}.
\ee
We fix the Hubble constant to $H_0=67.3\,\mrm{km}\,\mrm{s}^{-1}\mrm{Mpc}^{-1}$
and specialise to a spatially flat geometry, $\Omlo=1-\Ommo$.
The posterior is sampled using the affine-invariant MCMC sampler
\texttt{emcee} \cite{ForemanMackey:2012ig}.

\begin{figure}
\centering
\begin{subfigure}[b]{0.8\textwidth}
	\includegraphics[width=\textwidth]{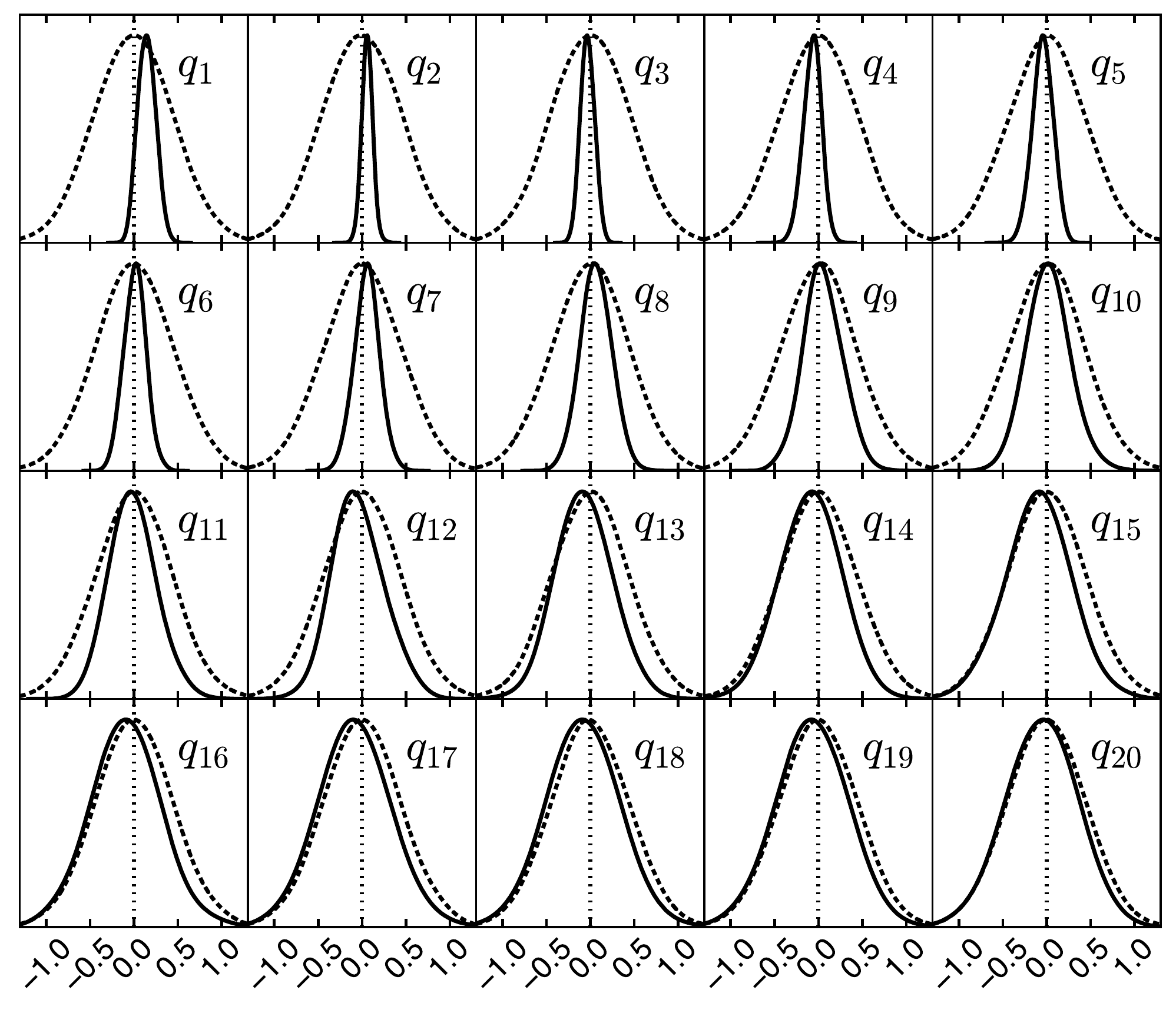}
\end{subfigure}
\begin{subfigure}[b]{0.8\textwidth}
	\includegraphics[width=\textwidth]{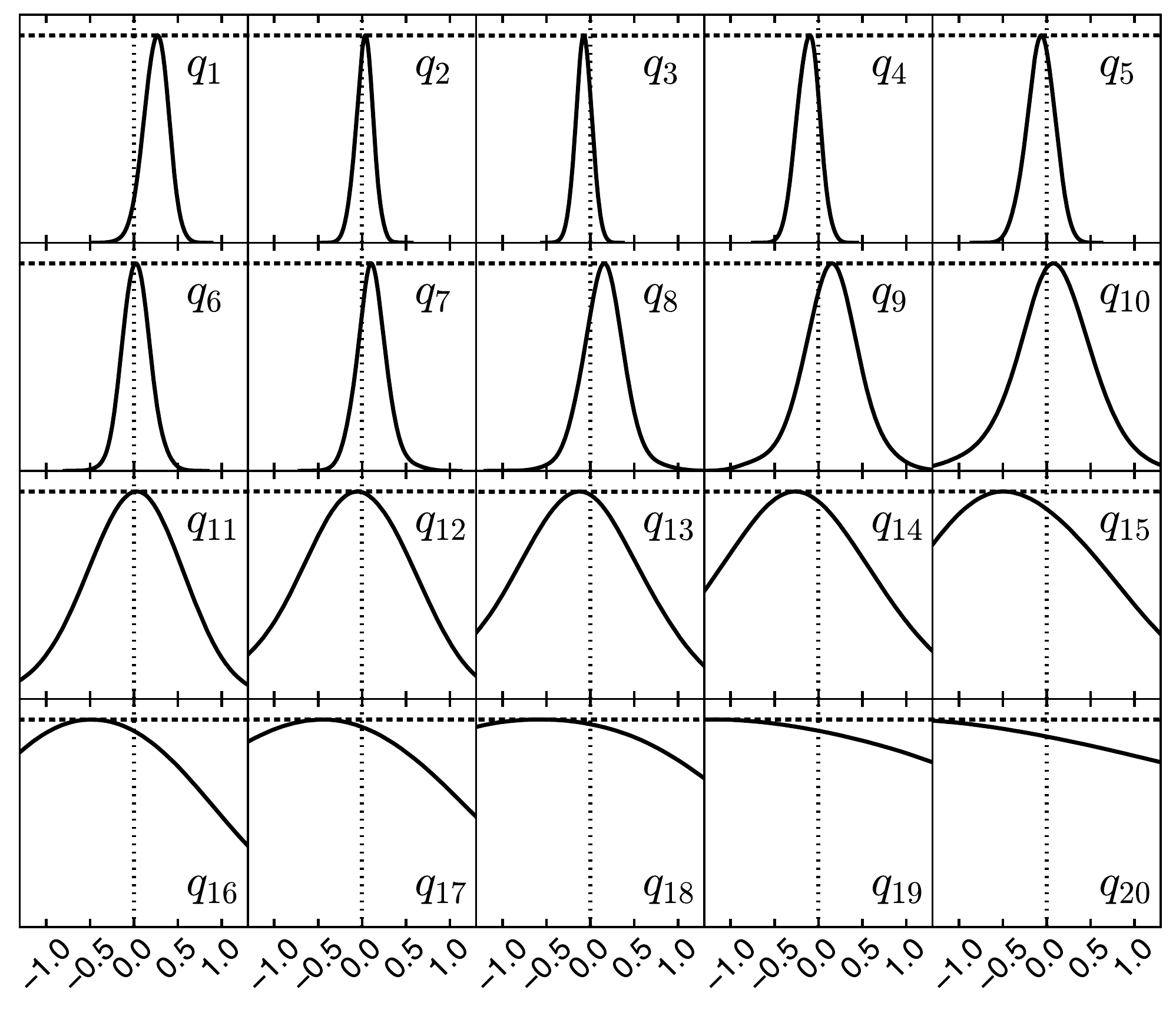}
\end{subfigure}
\caption{A comparison of the unnormalised 1-dimensional marginal posterior
		(solid curves) and prior (dashed curves) distributions
		for the 20 bin reconstruction with
		Prior I ({\em bottom panel}) and Prior II ({\em top panel}).}
  \label{fig:qmarg}
\end{figure}

\subsection{Reconstruction\label{sec:rec}}
The individual constraints for the 20 bin interaction model
are summarised in Figure \ref{fig:qmarg}, which are plotted together with the prior
probability distribution to show the improvement in each bin. It is clear the
first 10 amplitudes 
$q_1,q_2,\ldots,q_{10}$ covering a redshift range from $z=0$ to $z\approx0.4$
furnish the best constraints. The next few bins show mild improvements,
while the last five indicate very little constraining power at those redshifts.
As we have noted before this is due to these bins containing more, and better
quality, data than those at higher redshifts (and this is even after we used
logarithmically spaced bins in redshift space to account for the unevenly
distributed data). It should be noted that the CPZ prior requires bins of
uniform size (in redshift, scale factor, etc) to ensure $\cov_\pi$ is positive definite.
As Figure \ref{fig:qmarg} shows, Prior I is uninformative as to the amplitudes
size of the amplitudes and is essentially flat over an interval of moderate
values of $q_i$s.

In Figure \ref{fig:qrec} we reconstructed $q(a)$ (though shown as a function of redshift)
and show the probability density function (PDF) of $q(z)$. Although the constraints are not
strong it is clear that only for the first few $q_i$ are the limits tightest,
and in the case of Prior I the variance rapidly grows with redshift.
We recall that in the absence of any new information provided by the data
the joint posterior distribution is identical to the joint prior of $\bm{q}$.

We have checked that our reconstruction is robust to the number of bins by
changing to 10 and 30 bins --- both reproducing the basic features seen in
Figure \ref{fig:qrec}.

The reconstructed models will of course give a better fit to data than
flat \LCDM. To see just how much the fit improves we can compare the
$\chi^2\equiv-2\ln\calL(\theta;D)$, with lower values being preferred.
When evaluated at the mean parameter values we find a
$\Delta\chi^2\equiv\chi^2-\chi^2_\mrm{\Lambda CDM}=-2.2$ when using
Prior I and $\Delta\chi^2=-3.6$ when using Prior II. These represent 
very modest improvements in the quality of fit, given we have 20
more parameters than \LCDM. We present a model comparison in
Section \ref{sec:comparison}.

\subsubsection{How many modes?\label{eq:num-modes}}

\begin{figure}
	\centering
	\includegraphics[width=1\linewidth]{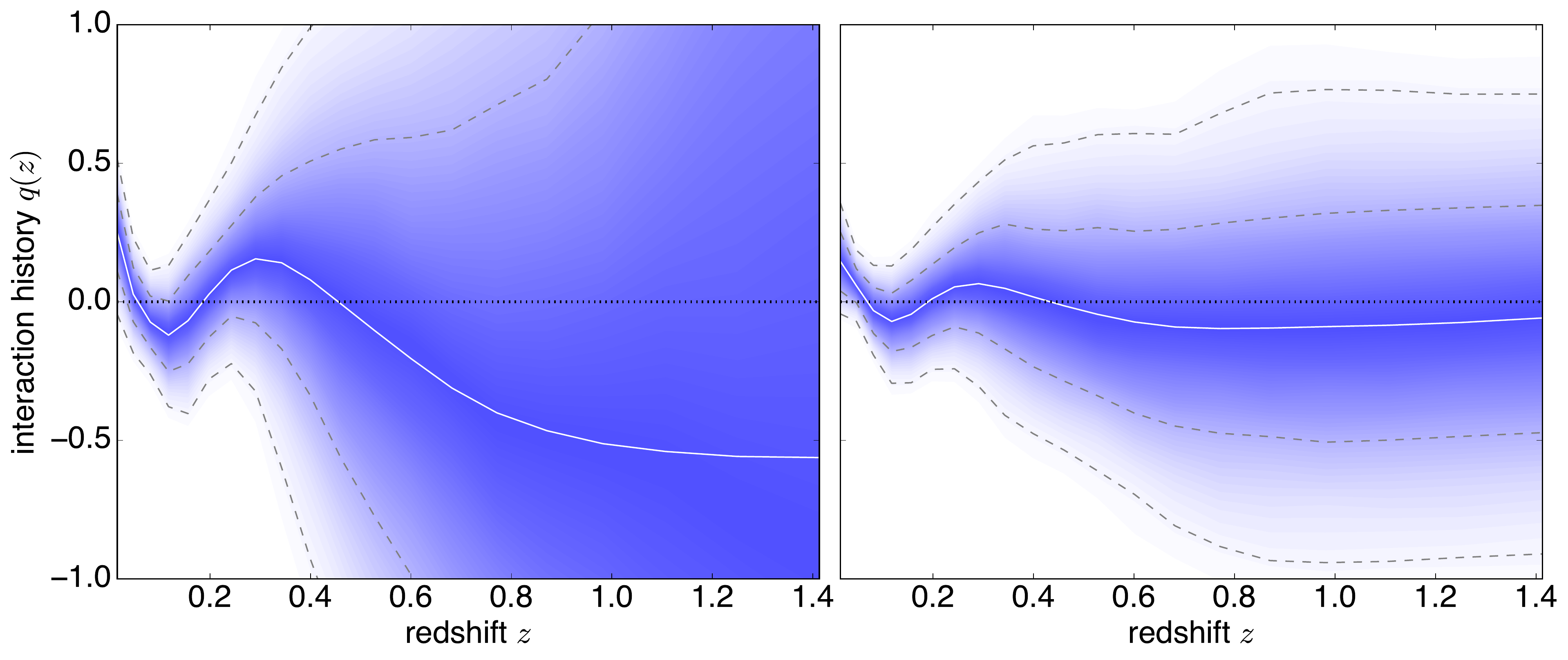}
\caption{The 20 bin reconstruction with Prior I ({\em left panel}) and
		Prior II ({\em right panel}). Dashed curves delimit the $68\%$
		and $95\%$ confidence ranges computed for each bin. The blue shading
		gives the PDF $p[q(z)]= p(q_1,q_2,\ldots,q_{20})$ and the mean
		$q_i$s are shown by the solid white curve.}
  \label{fig:qrec}
\end{figure}

\begin{figure}
	\centering
	\includegraphics[width=1\linewidth]{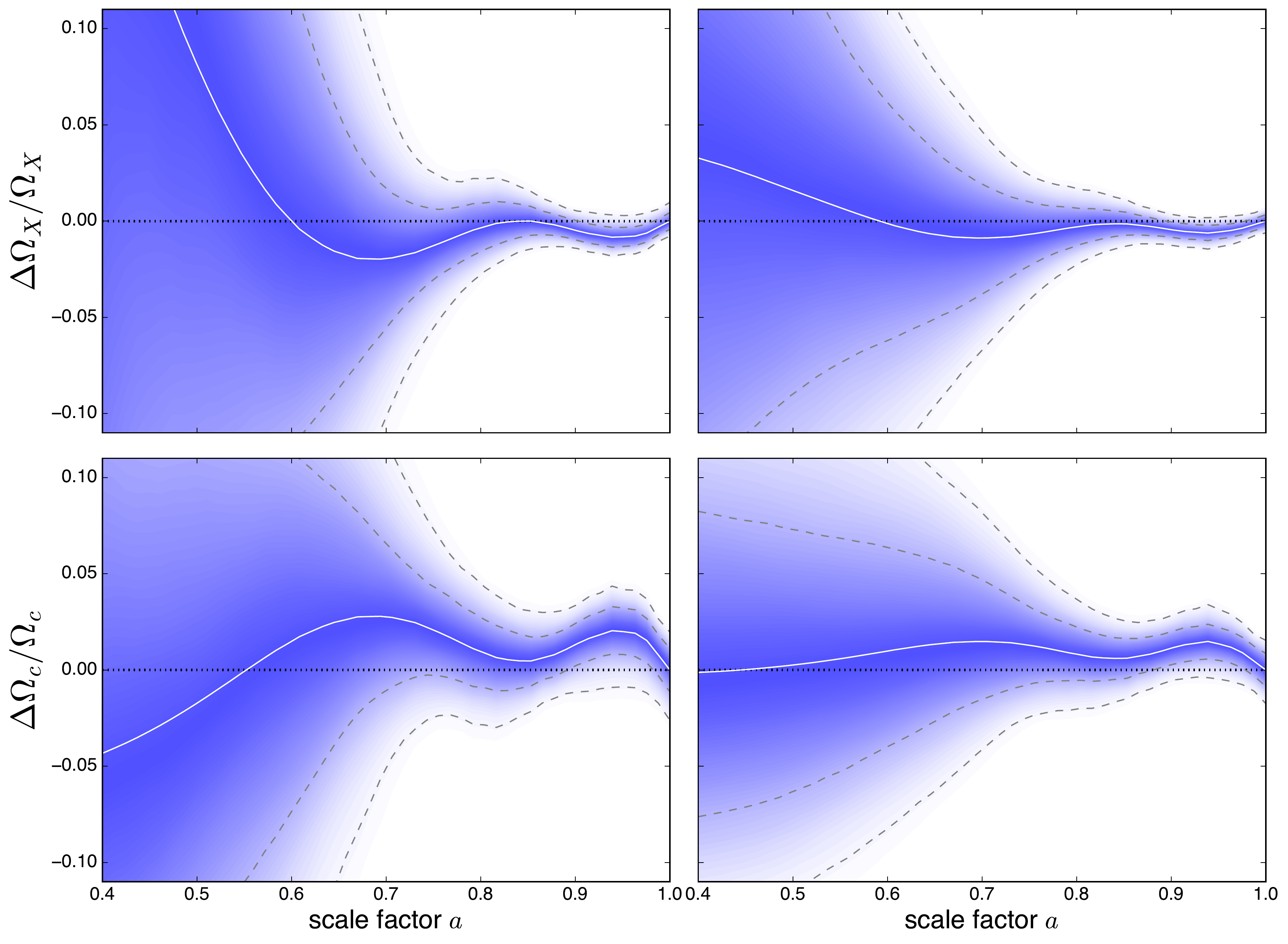}
\caption{The fractional change in the DM and DE density
		 relative to the non-interacting scenario
		 for the 20 bin reconstruction with Prior I
		 ({\em left panels}) and Prior II ({\em right panels}).
		 Here a positive change means a higher density
		 relative to \LCDM. The solid white curves shows the
		 evolution for the best-fit while the dashed curves indicate the
		 $68\%$ and $95\%$ confidence intervals.}
  \label{fig:delta-Omega}
\end{figure}

\begin{figure}
	\centering
	\includegraphics[width=1\linewidth]{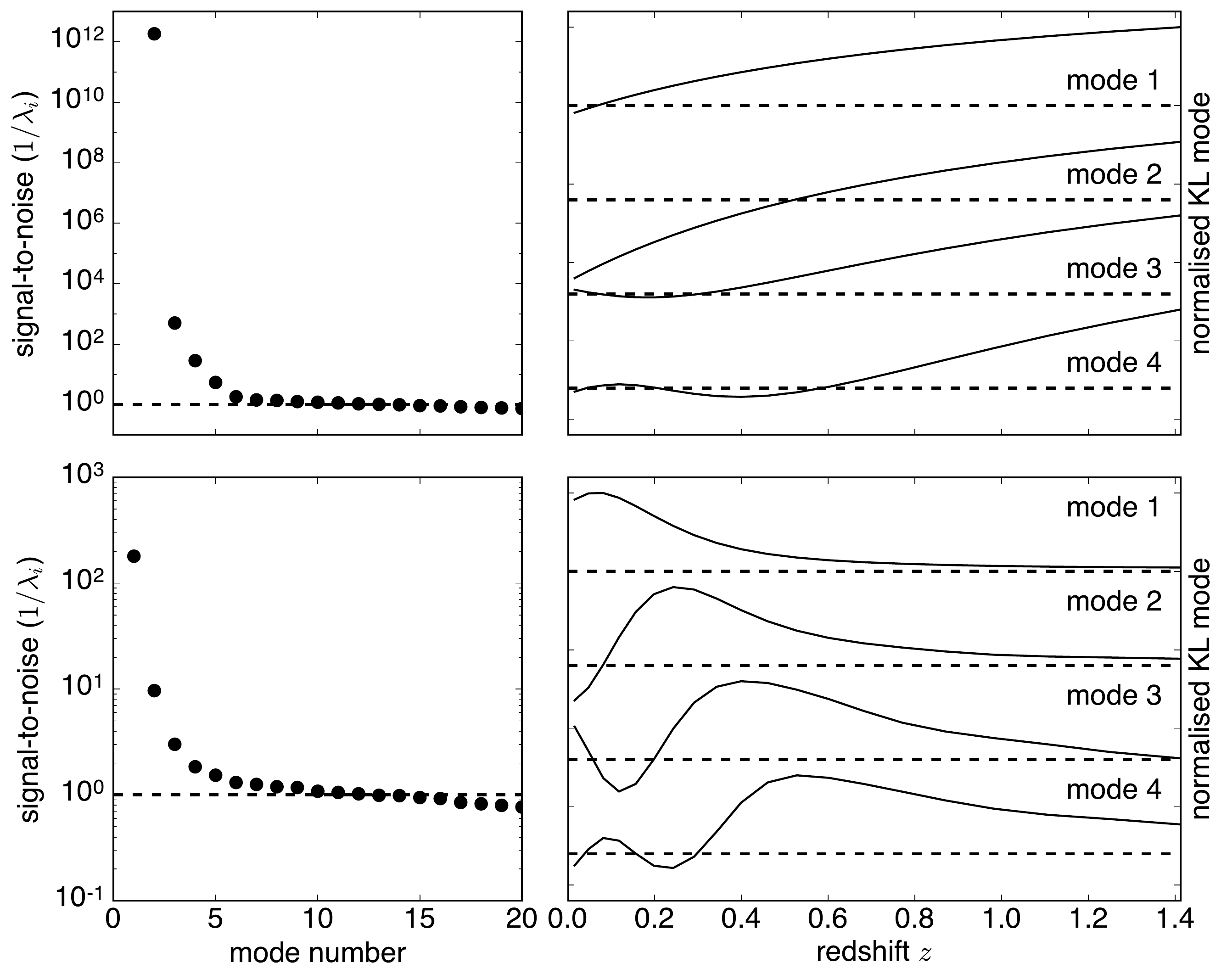}
	\caption{Karhunen-Lo\`{e}ve modes for Prior I (\emph{top panels})
			and Prior II (\emph{bottom panels}), both with $\xi_0=0.2$.
			In general the
			$i^\mrm{th}$ mode crosses zero $i-1$ times. The four best modes
			(\emph{right panels}) are normalised so that all have
			the same height. We offset each by a fixed amount with the dashed
			horizontal line indicating the zero point.}	
  \label{fig:KL}
\end{figure}

Having reconstructed $q(a)$ the question arises as to just how many KL
eigenmodes $m$ should be retained in the expansion \eqref{eq:q-KL}: too many
and we risk fitting the noise, while too few we fail to capture the physical
features being suggested by the data. To assess this
trade-off we compute the mean squared error given by
$\mrm{MSE}=\mrm{bias}^2+\mrm{variance}$ (see, e.g.\ \cite{Huterer:2002hy}).
Generally, the bias decreases with $m$, while
the variance increases with $m$. In our decorrelated basis it is given by
\be\label{eq:MSE}
\mrm{MSE}=\sum_{i=1}^n \left(q_i^{(m)} - \bar{q}_i\right)^2 
	+ \sum_{i=1}^n \left(\sigma^{(m)}_i\right)^2,
\ee
where
\be
\sigma^{(m)}_i = \bigg[\sum_{j=1}^m  e^2_j(a_i) \, \sigma^2_{\alpha_j}\bigg]^{1/2}
\ee
is the error of the reconstructed $q(a)$ in the $i^\mrm{th}$ bin keeping
only the first $m$ modes, each having a variance $\sigma^2_{\alpha_j}=1$
in the KL basis.
Moreover, $q_i^{(m)}$ is the associated mean of $\sigma^{(m)}_i$, whereas
$\bar{q}_i$ is the mean in the $i^\mrm{th}$ bin in the original
parametrisation (i.e.\ keeping all modes). A caveat to using \eqref{eq:MSE}
however is that we assume the true $q(a)$ is given by the mean of the full
reconstruction.
The optimal number of modes $m$ to keep is thus determined by minimising 
\eqref{eq:MSE}.
For both Priors I and II we find the MSE is minimised when $m=1$.

We next compute the Bayesian complexity $C$, using the Gaussian approximation
given by \eqref{eq:BC}.
Out of the 20 parameters we find $C=5.2$ for Prior I and
$C=3.2$ for Prior II.
We show in Figure \ref{fig:KL} the first four modes; above $i=5$ for
Prior I and $i=3$ for Prior II the modes generally become less smooth as we
expect.
The third PC in fact provides little information above $z=0.2$. It is clear that
the most interesting features
of $q(z)$ in our reconstruction is for low-redshifts. This could be for two
reasons. The first is that this could be a data effect caused by the specific
cosmological probes being used.
Unsurprisingly, these two describe the low-redshift features we see in Figure \ref{fig:qrec}.
For reference we list the three best constrained eigenmodes: (Prior I)
$\alpha_1=1.5\pm 1.0$, $\alpha_2=0.23\pm 1.00$ and $\alpha_3=0.48\pm 1.00$,
and (Prior II) $\alpha_1=1.2\pm 1.0$, $\alpha_2=0.49\pm 1.00$ and
$\alpha_3=0.89\pm 1.00$.
It is clear all except the first mode is consistent with zero (and the constraints
only get worse for larger $i$). However, the deviation of $\alpha_1$ from zero is
not statistically significant (at the $\sim 1\sigma$ level).

From the posterior $p(q_1,q_2,\ldots,q_{20})$ estimated from all MCMC samples
we can obtain the joint posterior of $\alpha_1,\alpha_2,\ldots,\alpha_m$ from
\be
p[q(a)] \equiv  p(q_1,q_2,\ldots,q_n) = p(\alpha_1,\alpha_2,\ldots,\alpha_n).
\ee
To do this we condition on the $q_i$s and marginalise over $\alpha_i$,
for $i=m+1,m+2,\ldots,n$, and use that the joint PDF of $\alpha_i$s
are separable:
\begin{align}
p\big[q^{(m)}(a)\big]
&=\int p(q_1,q_2,\ldots,q_n)\,
	 p(\alpha_1,\alpha_2,\ldots,\alpha_m\,|\,q_1,q_2,\ldots,q_n)\,
	 dq_1\,dq_2\ldots dq_n \nonumber \\
&= \int p(q_1,q_2,\ldots,q_n) 
	\prod_{i=1}^m\delta\big(\alpha_i-\sum_{j=1}^{n} A^{-T}_{ij}q_j\big) \,
	dq_1\,dq_2\ldots dq_n,
		\label{eq:p-qm}
\end{align}
where $\delta$ is the Dirac delta function.
In effect, we project $q(a)$ onto a subspace spanned by a subset
of the KL modes, which we achieve in practice by discarding modes $m+1$
and higher. Indeed, in the case $m=20$ we have $q^{(m)}(a)=q(a)$
so that we recover the results of Figure \ref{fig:qrec}.

\begin{figure}
	\centering
	\includegraphics[width=1\linewidth]{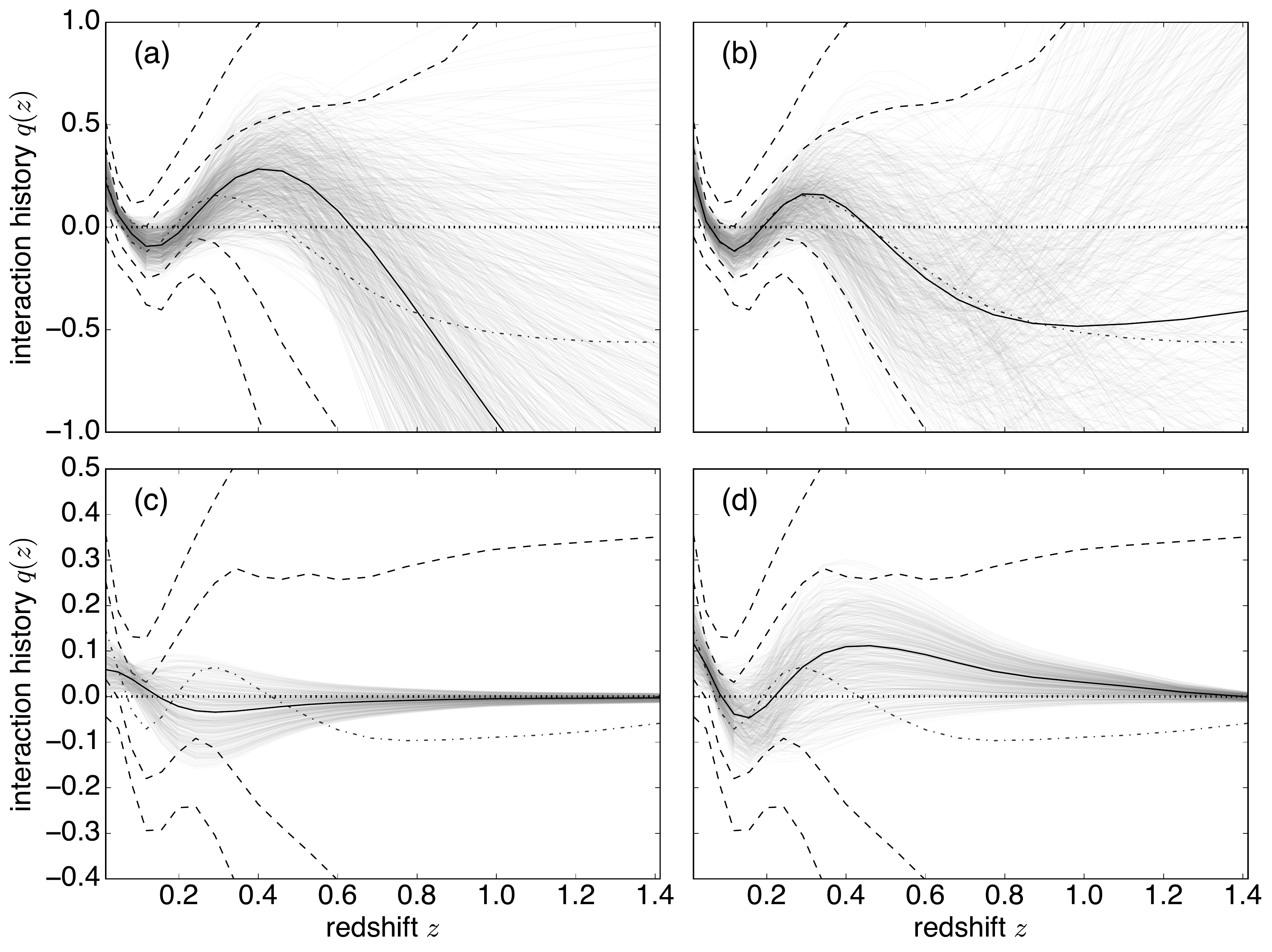}
\caption{\emph{Panel} (a): Reconstruction of $q^{(m)}(a)$, $m=4$ for 500 samples
		drawn from \eqref{eq:p-qm} lying within the $68\%$ confidence region.
		\emph{Panel} (b): Same as (a) but with $m=5$. \emph{Panels} (c), (d):
		Same as (a) and (b) but with Prior II and $m=2$ and $m=3$, respectively.
		The solid black curve in each panel indicates the truncation of \eqref{eq:q-KL}
		at the first $m$ KL modes. The dash dotted curve indicates the
		mean $q(a)$ (as shown in by the white curve in Figure \ref{fig:delta-Omega}).
		For comparison with Figure \ref{fig:qrec}, we indicate
		by dashed black curves the $68\%$ and $95\%$ confidence ranges of
		the full reconstruction.}
	\label{fig:qrec-KL}
\end{figure}

Figure \ref{fig:qrec-KL} shows the reconstruction $q^{(m)}(a)$ for the
leading $m$ KL modes, the leading $m-1$ KL modes, and also the PDF of the
reconstruction \eqref{eq:p-qm} for both priors.  This is the reconstruction
being constrained by the data, i.e.\ after eliminating the extraneous prior modes.
The number of modes $m$ are determined by the Bayesian complexity \eqref{eq:BC},
which tells us heuristically how many KL modes are being probed by data.
It can be seen that without the higher $m$ modes the variance at
redshift is significantly reduced. The fifth mode of Prior I, however,
dominates the variance above a redshift of $z=0.6$. As can be seen in
Figure \ref{fig:KL}, this mode has $S/N\gtrsim 1$ indicating this mode
ought to be discarded, too. Generally we find that the Bayesian
complexity provides a useful way to identify the relevant modes.

As expected, the features that the data are actually constraining (i.e.\ 
the low-redshift features) in Figure \ref{fig:qrec-KL} are preserved
after discarding the higher $m$ modes. As can be further seen, the fifth
mode of Prior I does not possess significant explanatory power as dropping
this still reproduces the basic form of the full reconstruction.
By constrast, the third mode of Prior I is crucial in reproducing the
low-$z$ feature ($z\lesssim 0.4$) of the full reconstruction.

As mentioned in Section \ref{sec:rec}, the improvement in $\chi^2\equiv-2\ln\calL$
for the full reconstruction gives $\Delta\chi^2\approx -3$. This is
because any deviations of the best-fit ${q}_i$ from zero are small, leading
to changes in the DM and DE density of less than $5\%$ at redshifts
$z\lesssim 0.6$. It is evident that subspaces of the full
reconstruction of $q(a)$ will only yield improvements to the $\chi^2$ of
$|\Delta\chi^2|\lesssim 3$. 
From the Akaike Information Criterion $\mrm{AIC}=\chi^2+2k$, for $k$ parameters,
we see that the introduction of an additional parameter must decrease the
$\chi^2$ by more than four to be competitive. We thus see that the
case for any non-zero $q_i$ is not strong.

\subsubsection{Sensitivity analysis}
In reconstructing $q(a)$ using \eqref{eq:xi} we have specified the tuning
parameter $\xi_0$ that sets the strength of the correlations. We have 
chosen $\xi_0=0.2$ for both Priors I and II, but different choices
are of course possible and it is worth exploring how our inference depends
on $\xi_0$. 
Rather than perform the analysis for a range of different $\xi_0$ we can
also marginalise over it. Thus we expand the hypothesis space to include
$\xi_0$ as a free parameter. This parameter is however unconstrained by data, as the 
joint likelihood does not depend on it. Nevertheless we can assign it a
prior $\pi(\xi_0)$ and fold it into the original prior \eqref{eq:pi-q}:
\be\label{eq:p-q-prior}
\pi(\bm{q}) 
= \int d\xi_0 \, \pi(\bm{q}|\xi_0) \, \pi(\xi_0),
\ee
where $\pi(\bm{q}|\xi_0)$ will be given by \eqref{eq:pi-q}.
As $\xi_0$ is a scale parameter we assign it a logarithmically
uniform distribution $\pi(\xi_0)\propto 1/\xi_0$.
For convenience we define the $\xi_0$-independent covariance
$\bar\cov_\pi\equiv\cov_\pi/\xi_0$, where $\cov_\pi$ can be either
the covariance of Prior I or II. In general, for $n$ interaction
parameters, by integrating $\xi_0$ over the range $[\xi_a,\xi_b]$ we obtain
\be\label{eq:p-q-int}
\pi(\bm{q})
\propto
	\frac{1}{\big(x^2(\bm{q})\big)^{n/2}}\:
	\bigg\{
		\Gamma\bigg(\frac{n}{2},\:\frac{1}{\xi_b}x^2(\bm{q})\bigg)
		-	\Gamma\bigg(\frac{n}{2},\:\frac{1}{\xi_a}x^2(\bm{q})\bigg)
	\bigg\},
\ee
where $x^2(\bm{q})\equiv(\bm{q}-\qfid)^\T \bar\cov^{-1}_\pi(\bm{q}-\qfid)/2$
and $\Gamma(s,x)$ is the incomplete gamma function. We consider a wide
interval with limits $\xi_a$ and $\xi_b$ that enclose $\xi_0=0.2$, as used in
\eqref{eq:pi-q}. Compared with the previous priors used, \eqref{eq:p-q-int}
has a heavier tail so that regions in parameter space far from $\qfid$ are
more easily explored. Moreover, since $\xi_a<0.2$ we now have a more narrowly
peaked mode at $\bm{q}=\qfid$, reflecting more confidence in the fiducial
model.
(In the case of Prior II, \eqref{eq:p-q-int} can be centred on $\bm{q}=0$
by defining a new $\cov_\pi$ that absorbs the shift.) 
If $\xi_b\gg \xi_a$, then \eqref{eq:p-q-int} has faster than Gaussian dropoff
since $\pi(\bm{q})\sim (\xi_b^{-1} x^2)^{-1} e^{-\xi_b^{-1} x^2}$,
as $x^2\to\infty$ (ignoring multiplicative constants), so that
$\pi(\bm{q})$ will be more strongly peaked than the Gaussian prior
\eqref{eq:pi-q}.

\begin{figure}
	\centering
	\includegraphics[width=1\linewidth]{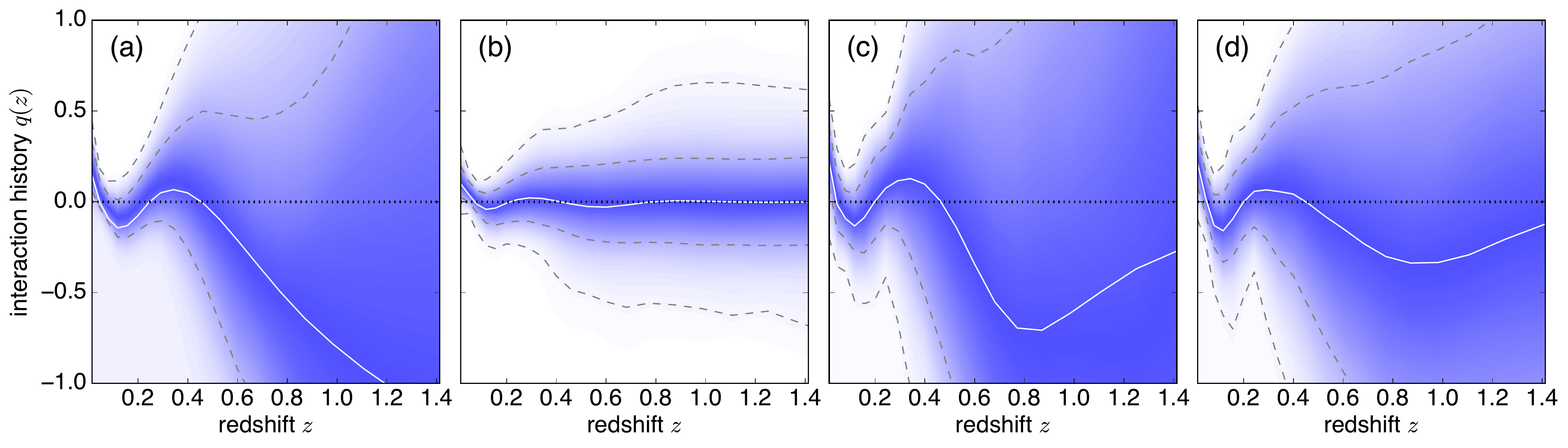}
	\caption{Same as Figure \ref{fig:qrec} but using: (a) prior \eqref{eq:p-q-int} with
			the running average $\qfid=\mat{R}\bm{q}$; (b) prior
			\eqref{eq:p-q-int} with $\qfid=0$; (c) Prior I with $\xi_0=2$; 
		    and (d) Prior II with $\xi_0=2$.}
  \label{fig:qrec-marge}
\end{figure}

Figure \ref{fig:qrec-marge} shows the full reconstruction using
the $\xi_0$ marginalised prior with bounds $\xi_a=0.02$
and $\xi_b=2$, i.e.\ enclosed by an interval with limits an order of
magnitude smaller or larger than $\xi_0=0.2$ used previously. 
Using $\qfid=0$ with \eqref{eq:p-q-int} we see this prior is more
constraining at high redshifts as the lower limit $\xi_a$ can be seen to
have the effect of causing $q(z)$ at redshifts $z\gtrsim0.2$ to hew more
closely to $q=q^\mrm{fid}$ and suppress the variance; c.f.\ Figure
\ref{fig:qrec}. The slight deviation for $z\lesssim 0.2$ is robust to
this prior, and also a more conservative choice of $\xi_0=2$ 
(i.e.\ $\xi_0$ ten times larger than before). Compared with Figure
\ref{fig:qrec}, Prior II shows larger deviations of the mean values of $q$
from $q=0$, particularly around $z\simeq 0.8$; this is due
to $q$ being less constrained to explore regions away from $\bm{q}=\qfid$,
but results in a larger variance around the mean. Where there are
large deviations (at higher redshifts) they are always accompanied by
substantial uncertainties related to the choice of prior.

\subsection{Simple one parameter extensions\label{sec:1par-ext}}
The previous sections have shown that while we are able to extract a handful
of modes with moderate to
high $S/N$, only the best mode provides a constraint on the associated
$\alpha_1$ that is not totally overwhelmed by its uncertainty. This indicates
the data are able to support at most one parameter.
The improvement in chi-square is marginal and we expect robust
model comparison with \LCDM\ to strongly disfavour the full reconstruction
model.

Instead of reanalysing the data using a subset of modes (which could
be considered using the data twice) we instead consider a one parameter model
covering a single wide bin spanning $z_\mrm{min}=0$ to $z_\mrm{max}=1.5$.
We will focus in particular on
constraints on the associated parameter $q$ and also ask whether the simplest
extension can be competitive with flat \LCDM.

For comparison with the ansatz \eqref{eq:Q} we will also consider
constraints from a physically motivated interaction given by
\be\label{eq:Q-gcg}
Q = qH\rho_X\left(1-\frac{\rho_X}{\rho_c+\rho_X}\right).
\ee
In contrast to \eqref{eq:Q}, which has an abrupt transition to non-interacting
\LCDM\ dynamics, this model smoothly interpolates in a logistic manner between
\LCDM\ at early times ($Q\to0$), to one with DM-DE interaction
($Q\sim qH\rho_X$; c.f.\ \eqref{eq:Q})
at late times.\footnote{The interacting model specified by \eqref{eq:Q-gcg} can
alternatively be viewed as one in which the interacting DM-DE is described
by a single fluid with an equation
of state $p=-A\rho^{-\alpha}$, where $A$ is a positive constant.
Such an exotic fluid is known as a generalised Chaplygin gas
\cite{Kamenshchik:2001cp,Gorini:2002kf}, and is notable for
having asymptotic behaviour that mimics CDM at early times and a
cosmological constant at late times \cite{Bento:2004uh,Makler:2002jv,Wang:2013qy}.}
We will call this model \logXCDM\ and the one bin model with \eqref{eq:Q} \IXCDM.
For both models we adopt flat priors on $q$, and again assume a spatially
flat background.

\begin{figure}
	\centering
	\includegraphics[width=\linewidth]{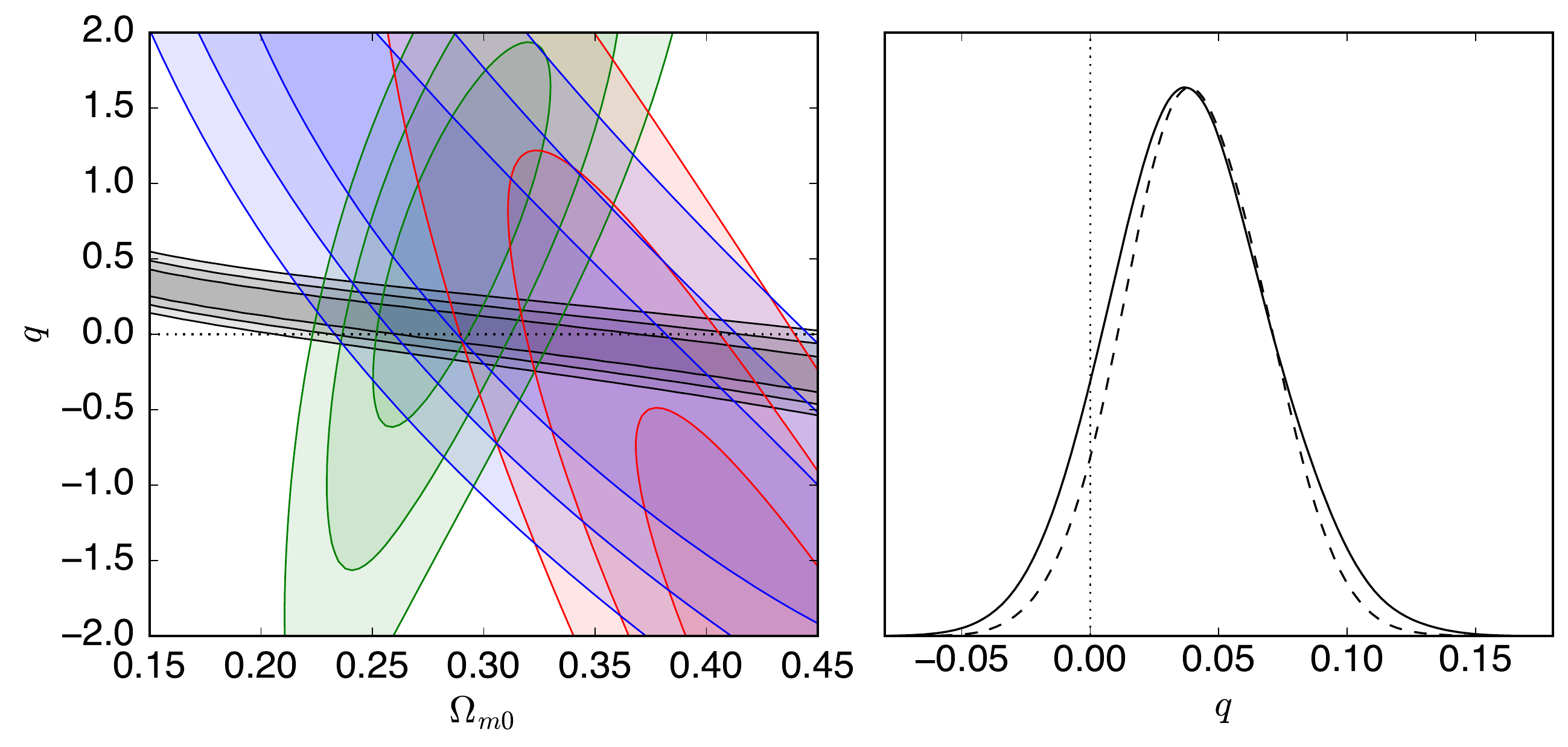}
\caption{\emph{Left panel}: Contours of the iso-likelihood 
		$-2\ln\mathcal{L}/\mathcal{L}_\mrm{max}$ equal to 2.3, 6.18, and 11.8
		in the $(\Ommo,q)$-plane for the \IXCDM\ model (fixing other
		parameters to the maximum likelihood estimate). The constraints are
		from each individual data set:
		RSD (black curves), BAO (green curves), SN (red curves), and
		cosmic chronometers (blue curves).
		\emph{Right panel}: 1D posterior of interaction parameter for
		\IXCDM\ (solid curve) and \logXCDM\ (dashed curve).}
  \label{fig:om-q-degen}
\end{figure}

Combining all data we find $q=0.039\pm 0.031$ for \IXCDM\ and a
slightly stronger constraint of $q=0.041\pm 0.027$ for \logXCDM, though
both are consistent with no interaction at $95\%$ confidence level.
The datum $\omb$ of the compressed CMB data set represents a highly
informative prior. Leaving this out of the data set we find a
much weaker constraint of $q=0.021\pm 0.029$, a $\approx 85\%$ shift in
the mean value.

Figure \ref{fig:om-q-degen} shows the constraints from individual
data sets obtained from their respective likelihoods. 
It is clear that individual data sets are not sensitive probes to $q$, except
for RSD data, which however shows a strong degeneracy with $\Ommo$.
It is therefore necessary to exploit the complementarity of data sets to
break the degeneracy. A positive $q$ yields relatively more matter and less DE
relative to \LCDM. The matter-radiation equality, on which the 
sound horizon depends, will occur earlier.
As the window between the epoch of matter-radiation
equality and the time of decoupling is wider, density fluctuations have 
more time to decay in the radiation-dominated epoch and the overall effect
being a suppression of the baryon acoustic peaks \cite{Valiviita:2009nu}.

Since the CMB data depends on an assumed cosmological model, it is
interesting to consider constraints on $q$ from local cosmological probes
only (BAO, CC, RSD, SN), which have the virtue of being model independent.
Though CMB data comprises three precise measurements, we find by leaving it
out we recover significantly worse constraints on $q$. In addition
cosmic chronometers data, while a direct measurement of the expansion
history, provide only slight improvements in constraints owing to the large
systematic errors and the fact that SN probes roughly the same physics
(see Figure \ref{fig:om-q-degen}).

\begin{figure}
	\centering
	\includegraphics[width=0.7\linewidth]{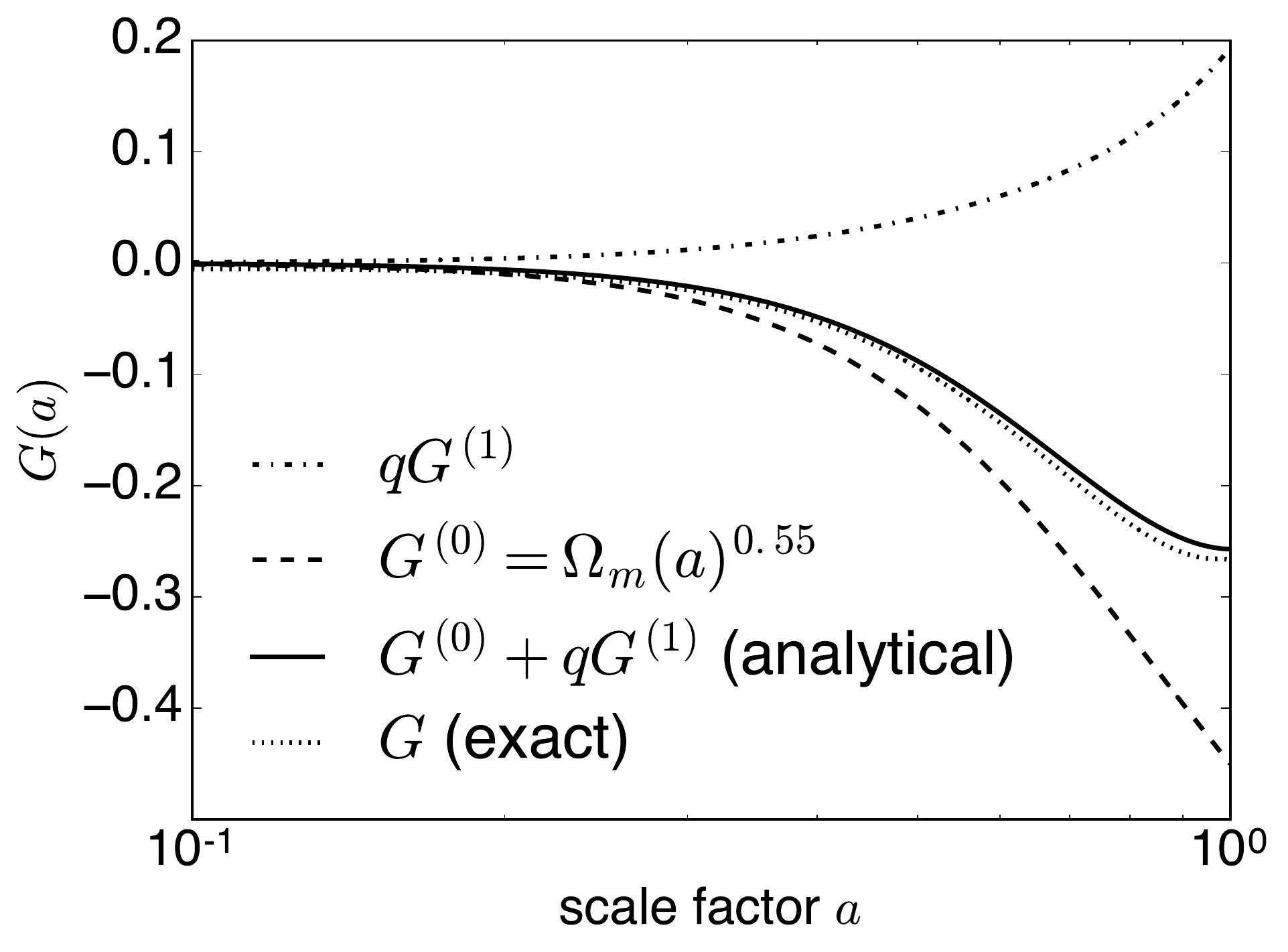}
\caption{Plot of $G(a)\equiv d\ln(D/a)/d\ln a=f(a)-1$, i.e.\
		the growth function with the matter mode removed 
		($G=0$ in a matter only universe).
		Shown is the exact numerical solution of
		\eqref{eq:dG} (dotted curve) and the linear
		approximation $G\approx G^{(0)}+qG^{(1)}$ (solid curve)
		with $q=0.1$, $\Ommo=0.3$, $\Omlo=0.7$ and $\gamma=0.55$
		for \logXCDM.}
		\label{fig:G}
\end{figure}

The origin of the dark degeneracy was discussed in \cite{Kunz:2007rk}.
The expansion history as probed by BAO and SN is sensitive only to the
evolution of the total energy content, which determines $H$ through
Friedmann's equation, and not to couplings between individual
components. The same evolution can be produced in more than one way. For
instance, a coupled dark sector is observationally indistinguishable from
one that is uncoupled with DE having a certain equation of state.
The interaction parameters only enter through the derivative of $T_{\mu\nu}$
and not through Einstein's equation. As a result the degeneracy persists
regardless of whether we consider background cosmological observables
or perturbation observables \cite{Kunz:2007rk}.
Although RSD places narrow constraints it comes with an almost exact
degeneracy with $\Ommo$ as seen in Figure \ref{fig:om-q-degen}.
We can see this quantitatively as follows. Noting that $q$ is small (of
order $10^{-2}$) we linearise \eqref{eq:growth-eq} about $q=0$ and solve
perturbatively. To do this we first reformulate \eqref{eq:growth-eq} in
terms of $f=d\ln D/d\ln a$.
We can thus recast \eqref{eq:growth-eq} as
\be\label{eq:dG}
\frac{df}{d\ln a}=f-f^2-p(a)f-r(a),
\ee
where $p(a)$ and $r(a)$ are given by the coefficients of the second and
third term of \eqref{eq:growth-eq}, respectively.
We write the interacting solution as the sum of non-interacting \LCDM\
component $f^{(0)}(a)$ and the interacting component $f^{(1)}(a)$:
$f(a)\approx f^{(0)}(a) + q f^{(1)}(a)$.
The zeroth-order solution is $f^{(0)}(a)=\Omm(a)^\gamma$, with the matter
density obeying the usual scaling and the growth index $\gamma$ equal 
to 0.55 for flat \LCDM\ \cite{Linder:2005in}.
Inserting this into \eqref{eq:dG} then formally integrating we find
\be\label{eq:G1}
f^{(1)}(a)
=-\frac{1}{u(a)} \int^a_0d\ln a'\,  u(a')
	\big[r^{(1)}(a')+p^{(1)}(a')\Omm(a')^\gamma\big],
\ee
where $u(a)=a^4H(a)$, $r^{(1)}(a)$ is the linear term of
$r\approx r^{(0)}+qr^{(1)}$ and likewise for $p^{(1)}$. 
As in \cite{Linder:2007hg} we have discarded the quadratic term
$(f^{(0)})^2$, and here additionally
$f^{(0)}f^{(1)}$. In the case of $Q$ given by \eqref{eq:Q-gcg} we have
$f^{(1)}\sim a^4$, as $a\to 0$, i.e.\ like \LCDM\, $G$ is small in the
early matter-dominated era.
Thus we can understand this degeneracy from Figure \ref{fig:G}:
less matter implies a greater abundance of DE which suppresses
the growth of structure. However, this can be offset by a positive interaction
rate ($q>0$), which raises the matter abundance across all epochs compared to
\LCDM\ (for the  same $\Ommo$) --- this explains the anticorrelation
between $q$ and $\Ommo$ as seen in Figure \ref{fig:om-q-degen}.

We can also compare how the growth function changes in the presence of interaction.
Recall in flat \LCDM\ the growth function $f(a)$ is completely
specified by the expansion history $H(a)$ (assuming also $\rho_m\propto a^{-3}$).
With interaction, $Q(a)$ alters $H(a)$ and $\rho_m(a)$, but also 
modifies the growth equation through additional terms. We can connect
\eqref{eq:G1} to the growth index formalism \cite{Linder:2005in,Linder:2007hg}
by integrating $f=d\ln D/d\ln a$ to obtain
\be
g(a)\equiv D(a)/a 
= A(a;q)\exp\left\{\int^a_0 d\ln a'\big[\Omega_m(a')^\gamma-1\big]\right\},
\ee
where we defined $A(a;q)=e^{q\int^a_0 d\ln a' f^{(1)}(a')}$ (since $f^{(1)}\sim a^4$
as $a\to 0$ this prefactor tends to unity as $a\to 0$; c.f.\ Figure \ref{fig:G}).
It is clear that the growth factor depends on $q$ in addition to the
growth index parameter $\gamma$. Thus if we attempt to fit $\gamma$
assuming $f=\Omm(a)^\gamma$ we are liable to find a systematic bias,
resulting in a shift of $\gamma$ from its \LCDM\ value of $0.55$.
The growth index formalism is commonly used to probe modified gravity but
care must be taken when interpreting $\gamma$ as this analysis shows it is
possible to have both a scale-indepedent growth factor $D(a)$
and a value of $\gamma\neq0.55$, and still have the theory of
gravity be given by general relativity.

\subsubsection{Model comparison\label{sec:comparison}}

Any model that is an extension of \LCDM\ is guaranteed to give an
equally good or better fit to data.
The question that must be asked of the model is whether the improvement
in fit justifies the additional flexibility. To address this we 
compute the Akaike Information Criterion (AIC) and the Bayesian
Information Criterion (BIC). For a model with $k$ parameters and
maximum likelihood $\calL_\mrm{max}$ the AIC is given by
$\mrm{AIC} = -2\ln\calL_\mrm{max} + 2k$, and the BIC is given by
$\mrm{BIC}=-2\ln\calL_\mrm{max}+k\ln N$, with $N$ the size of the data set.
It should be noted that while the AIC and BIC are very similar, differing
only in their penalty term, they are obtained from different starting points:
the AIC has its origins in information theory and is based on an approximation
of the Kullback-Leibler divergence (c.f.\ \eqref{eq:DKL}), while the BIC
estimates the (logarithm) Bayesian evidence under
the assumption that the likelihood is Gaussian and $N$ is large. The quantity
of interest however is the difference, and we will take positive values of
$\Delta\mrm{AIC}$ and $\Delta\mrm{BIC}$ to indicate preference for the
interacting model.

We find $\Delta\mrm{AIC}=-0.69$ when comparing spatially flat \IXCDM\ and \LCDM\ with
$k=12$ and 11 parameters, respectively. This indicates a slight preference for
\LCDM. We find also $\Delta\mrm{BIC}=-5.4$, indicating strong preference
for \LCDM\ when assessed on Jeffreys's scale.
In the case of spatially flat \logXCDM\ and \LCDM, we find 
$\Delta\mrm{AIC}=0.22$ and $\Delta\mrm{BIC}=-4.5$. Again the BIC
is decisively in favour of the non-interacting scenario.
That the BIC is more penalising than the AIC is not surprising: the AIC is
generally more accommodating of additional parameters when $N$ is large, but unlike
the BIC it is dimensionally inconsistent, in that the tendency of AIC to
select the more complex model does not fall as the size of the data set grows \cite{Liddle:2004nh}.

As a further comparison, we will also compare \LCDM\ with
the previous reconstruction of $q(a)$. While this may be seen as post-hoc
tuning of the interacting model, it is nevertheless interesting to see
whether this model can be competitive under favourable circumstances. We will
thus focus on a minimal model that keeps just the best measured mode, parametrised
by $\alpha_1$. This model is a one parameter extension of \LCDM\ upon
marginalising over all other $\alpha_i$. The model selection measure of
choice we take to be the Bayesian evidence
$E=\int d\theta\,\calL(\theta;D)\pi(\theta)$, which readily embodies
Occam's razor. Given two models, model selection is decided by 
computing the Bayes factor $B$ given by the ratio of the evidences.
We will take $\ln B<0$ to mean preference for the reconstructed model
and $\ln B>0$ to mean preference for \LCDM.
Since non-interacting models are nested within interacting models, we can
compute the Bayes factor using the Savage-Dickey density ratio formula
$B = {p(\alpha_1=0|D)}/{\pi(\alpha_1=0)}$. We recall in the KL basis 
$\alpha_1,\alpha_2,\ldots,\alpha_{20}$ are uncorrelated variables and 
the joint prior is also a Gaussian that can therefore be separated:
$\pi(\bm{\alpha})=\pi(\alpha_1)\pi(\alpha_2)\ldots\pi(\alpha_{20})$.
Marginalising over $\alpha_2,\alpha_3,\ldots,\alpha_{20}$ is trivial for
Gaussian distributions and we have that in this decorrelated KL basis
$\pi(\alpha_1)$ is given by a 1-dimensional Gaussian with a mean of zero
and a variance of $1/\lambda_1$ (for both Priors I and II).
Parameters shared by both models will also have the same priors.
We find $\ln B=-1.4$ for Prior I and $\ln B = -1.9$ for Prior II,
indicating weak to moderate evidence for \LCDM\ as judged on Jeffreys's scale,
although not as decisively as in the one parameter models considered earlier.
The slightly less favourable evidence in the case of Prior II is to be
expected as this prior takes $\qfid=0$, which necessarily drives the
regression towards reconstructions consistent with null interaction.

\section{Future prospects: the Fisher forecast\label{sec:forecast}}
As is clear from the reconstruction we are not yet able to obtain tight
constraints on a possible DM-DE coupling.
A more detailed reconstruction of the finer features of $q(a)$ may be
possible in the future with upcoming stage-IV surveys, such as with the Large Synoptic
Survey Telescope (LSST) \cite{Abell:2009aa} and the Dark Energy Spectroscopic
Instrument (DESI) \cite{Aghamousa:2016zmz}.
In anticipation of this we use the Fisher framework to forecast improvements in
the constraints obtained in this analysis.

We assume a SN redshift distribution of the form
\be
\frac{dN}{dz} \propto z^2 \, e^{-(z/z_0)^\alpha}.
\ee
For the LSST survey we take $z_0=0.04$ and $\alpha=0.7$.
As LSST is expected to deliver $>10^4$ SNe Ia per year for ten years
we consider what constraints we might obtain with a one year sample with a
total number of $N_\mrm{tot}=5\times 10^4$ SNe Ia
and the full ten year sample with $N_\mrm{tot}=5\times 10^5$ SNe Ia.
We thus draw $N_\mrm{tot}$ samples distributed according to the probability
density function $p(z)=(1/N_\mrm{tot})dN/dz$.
For simplicity we adopt the conventional SN analysis in which the 
data are Gaussian distributed and the Fisher matrix is given by
\be
F_{ij}^\mrm{LSST-SN}
=\sum_{k=1}^{N_\mrm{tot}} \frac{1}{\sigma_{m_k}^2}
	\frac{\partial\mu}{\partial\theta_i}\frac{\partial\mu}{\partial\theta_j}\bigg|_{z=z_k},
\ee
where the indices $i$ and $j$ runs over all parameters.
We take the total error of the apparent magnitude $\sigma_{m_k}$
to be made up of an intrinsic scatter $\sigma_\mrm{int}$ and
assume a Gaussian redshift uncertainty $\sigma_z$. Propagating the redshift
error by \eqref{eq:distmod} the magnitudes have a total squared error given by
$$
\sigma^2_{m_k}=\sigma_\mrm{int}^2+\left(\frac{5}{z_k\ln 10}\right)^2 \sigma_{z_k}^2.
$$
Here $\sigma_\mrm{int}=0.12$ and we assume photometric redshift
errors modelled by a linear drift, $\sigma_{z_k}=0.05(1+z_k)$.

\begin{figure}
	\centering
	\includegraphics[width=0.65\linewidth]{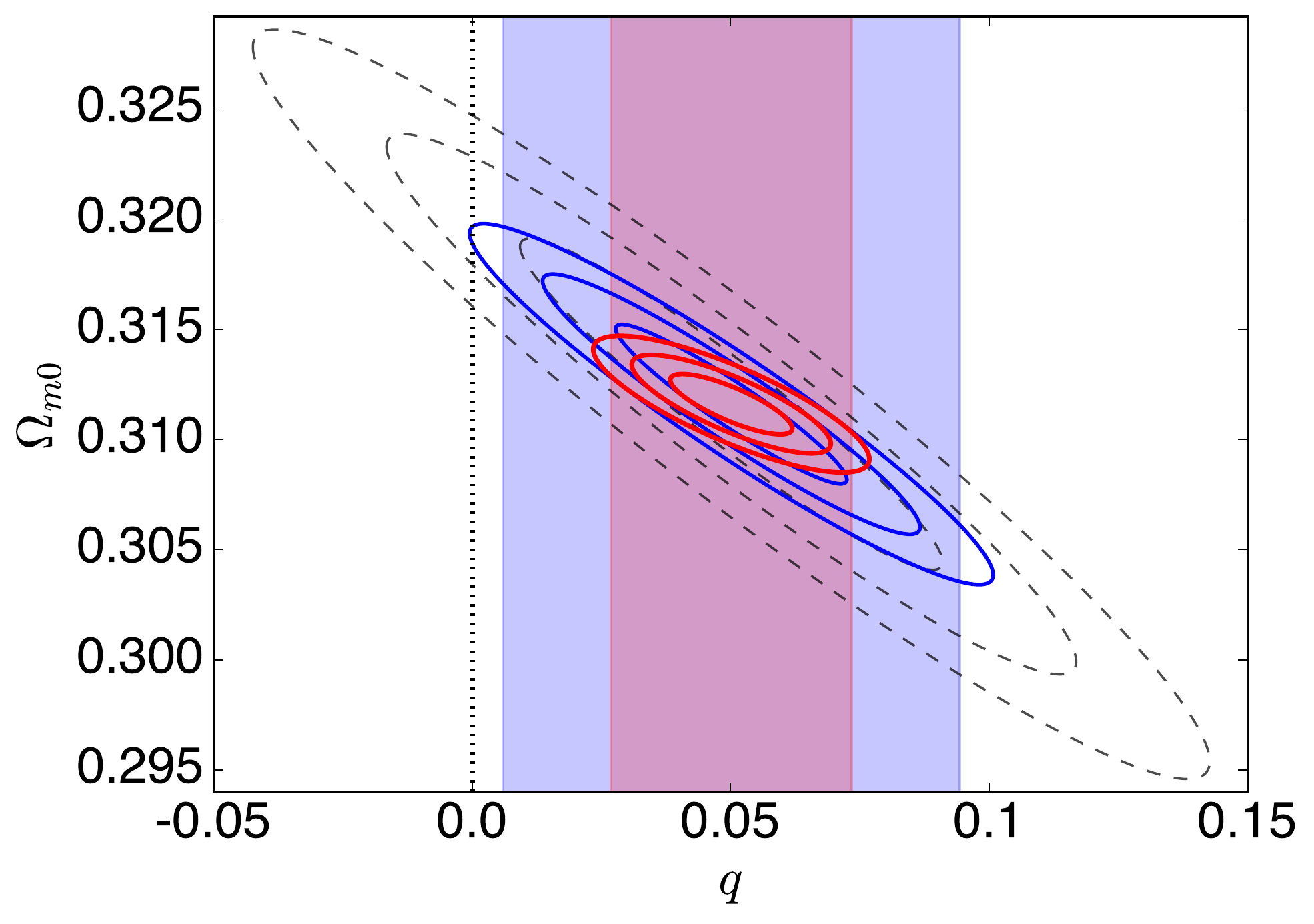}
	\caption{The projected $68\%$, $95\%$ and $99\%$ confidence regions
			for $5\times10^4$ (solid blue contours) and $5\times10^5$
			SNe Ia (solid red contours), adopting reference values of 
			$\Ommo=0.3116$ and $q=0.041$. Also shown are the $99\%$ confidence
			intervals for $q$ using $5\times10^4$ (blue band) and $5\times10^5$ (red band)
			SNe Ia. The dashed black curves show constraints obtained
			in this analysis.}
	\label{fig:forecast}
\end{figure}

In addition to SN constraints from LSST we also include constraints from RSD
data from DESI and the DESI Bright Galaxy survey \cite{Aghamousa:2016zmz}.
We use the projected RSD constraints based on a sky coverage of $14,000\,\text{deg}^2$
and the pessimistic wavenumber cutoff of $k_\mrm{max}=0.1h\,\text{Mpc}^{-1}$.
The DESI estimates are of $f\sigma_8$ at 18 redshifts between $z=0.05$ and
$z=1.85$ (see Tables 2.3 and 2.5 in \cite{Aghamousa:2016zmz}). We
further supplement this with DESI projected errors on the transverse and
radial BAO scales, $d_A/r_s$ and $Hr_s$, respectively, and which are
correlated measurements with a correlation coefficient of 0.4.
We assume negligible correlations between current and future galaxy
surveys, i.e.\ we assume the data are not being double counted.

We forecast constraints using the Fisher information matrix
\be\nonumber
F_{ij}
= F_{ij}^\mrm{LSST-SN} + F_{ij}^\mrm{DESI-RSD} + F_{ij}^\mrm{DESI-BAO}
		+ F_{ij}^\mrm{data},
\ee
where $F_{ij}^\mrm{data}$ is computed from the joint likelihood and
represents hypothetical future prior information obtained in this work
from current data. To obtain a sense of the improvement in constraints
possible we focus on a single wide bin between $z=0$ and $z=1.5$.
We evaluate the Fisher matrix at the maximum likelihood
estimate using \logXCDM, which we find to be $\Ommo=0.3116$,
$\Ombo=0.04977$, $q=0.041$, and $\sigma_{80}=0.767$.
As $q$ is most strongly correlated with $\Ommo$ we show in Figure
\ref{fig:forecast} the joint constraints in the $(q,\Ommo$)-plane
having marginalised over all other parameters.
We find a constraint of $\Delta q=\pm0.019$ and $\pm0.0078$
($68\%$ C.L.) for the one and ten year SN sample, respectively.
For the full ten year LSST observing run we expect the constraints on
$q$ to improve by a factor of $\approx 3.5$. Although this represents
only a $\approx 20\%$ determination of $q$ it is still sufficient to rule
out the no-interaction hypothesis at $99\%$ confidence level.

\section{Conclusions\label{sec:concl}}
In this work we have investigated in a model-independent way a range of
minimal extensions to \LCDM\ that relax the usual assumption that DM and
DE are non-interacting species.
We adopted a non-parametric approach and reconstructed directly from data
the interaction history within a low-redshift window of $z<1.5$.
Using a range of primarily low-redshift cosmological probes
we do not see statistically significant evidence for coupling between
DM and DE, although we note that
the tightest constraints obtained in the reconstruction are for $z\lesssim 0.4$, 
which also happen to show the strongest signs of a departure from \LCDM.
While these departures are $\lesssim 2\sigma$ in statistical
significance, they suggest a very mild late-time breakdown of the assumptions
of \LCDM.

We find that although several Karhunen-Lo\`{e}ve modes of the reconstruction can be
cleanly extracted from the data, only one mode can be satisfactorily constrained.
That most of the interaction parameters have substantial uncertainties is not surprising
given the weak constraints on $q(a)$ can be likened to that of the dark
energy equation of state $w_X(a)$. Current data are not yet
able to rule out the cosmological constant scenario using
the CPL parametrisation \cite{Chevallier:2000qy,Linder:2002et} 
$w_X(a)=w_0+w_a(1-a)$, which is the simplest dynamical dark energy model. 
Like $w_X$ the interaction parameter $q$ directly modify the scaling relations
of the energy density and so enter $\rho_X$ and $\rho_c$ through an integral.
On the other hand geometric probes are built from distance measures that are
integrals over the expansion and thus sensitive to $q$ and $w_X$
through a double integral. In both cases this limits the effectiveness of such
probes and emphasizes the need for dynamical probes, such as RSDs.
However, in the case of interaction it is also necessary to combine probes
to break the `dark' degeneracy to obtain tight constraints \cite{Kunz:2007rk}.
Therefore the situation with interaction should not be expected to yield
strong constraints even for the minimal models considered here.

The added flexibility of an extra degree of freedom should always be
weighed against the increased model complexity, and in this work we have
also assessed the viability of interacting models. We find that \LCDM\
remains the preferred model over all interacting scenarios considered
here. Even under favourable conditions, the most competitive interacting
model of the reconstructions (using the best constrained KL mode only) we find
that \LCDM\ is still favoured but with weak Bayesian evidence ($|\ln B|=1.4$).

Our constraints on $q$ for the minimal models (Section \ref{sec:1par-ext}) are
consistent with those in the recent literature using ansatz forms of $Q$.
In \cite{Yang:2016evp} four parametrisations of $Q$ were investigated, finding
interaction to be statistically insignificant in each.
More recently, in \cite{Yang:2018euj}, a Bayesian comparison is performed
using two novel parametrisations of $Q$ proportional to the
dark matter density that specially overcome the early-time instability \cite{Valiviita:2008iv}.
For various data combinations, the authors find that \LCDM\ is always favoured
with Bayesian evidence ranging from ``positive'' to ``very strong''.
Similarly in \cite{DiValentino:2017iww}, interaction is investigated in the
context of relieving the $\HO$ tension. The authors find moderate Bayesian
evidence $\ln B=1.8$ for interaction using Planck 2015 and a prior on $\HO$ given by
the SH0ES 2016 determination \cite{Riess:2016jrr}. However, by also adding
BAO and SN Ia data the Bayes factor reduces to $\ln B=0.53$, i.e.\ inconclusive
evidence for interaction.
In \cite{An:2018vzw} a more elaborate quintessence-type interaction
model is studied, in which a spin-$\frac{1}{2}$ fermionic field (DM)
is coupled to a canonical scalar field (DE) by the Yukawa interaction, following
the earlier work of \cite{Costa:2014pba}. This model is described by two additional parameters for
the Yukawa potential ($\lambda$) and the coupling ($r$). Though KiDS weak lensing
and CMB lensing mesaurements are considered, the tightest constraints are found
using only the Planck 2015 data set (TT, TE, and EE):
$r=-0.01074^{+0.0424}_{-0.0426}$ at $68\%$ confidence level, with $\lambda$
found to be poorly constrained. (Interestingly, using only the Planck
data set, the authors find this model is moderately favoured over \LCDM\ using
the Deviance Information Criterion.) 

The notable exception to these results reaffirming \LCDM\ is that of 
Salvatelli et al.\ \cite{Salvatelli:2014zta}, who reported a late-time
interaction that excluded \LCDM\ at $99\%$ confidence level. While these
results generated much interest in the phenomenology of interacting models
as viable extensions to \LCDM, but to date the results of \cite{Salvatelli:2014zta}
have not been borne out by recent data.
In particular, the recent work of \cite{Martinelli:2019dau} reanalysed
the same interacting models using recent data sets and found no significant
deviation from \LCDM. They suggest that the differences in results with the
earlier analysis are due simply to differences in the Planck 2013 and 2015
data sets, with the latter no longer favouring interaction. However, we also note
that the constraints of \cite{Martinelli:2019dau} on the same model are more
stringent, coming from a combined analysis of CMB and the low redshift probes
of RSD, BAO, and SN. By contrast the results of \cite{Salvatelli:2014zta} are
based on CMB (Planck 2013 temperature fluctuations and WMAP polarization) and
RSD data.
While interacting scenarios can resolve tensions in local analyses
\cite{DiValentino:2017iww,Yang:2018euj}, \LCDM\ is still robust against a wide
variety of probes.

Though we have aimed to be as model-independent as possible, we have
nevertheless had to make assumptions about the type of
interaction. We have thus assumed the covariant interaction 4-vector is 
directed along the geodesic flow of CDM (i.e.\ no momentum
transfer) and also that DE has an equation of state $w=-1$. With these
assumptions we introduce no additional dynamical degrees of
freedom and maintain the picture of DE as a vacuum energy, only now with
the possibility of an energy exchange with CDM.
Further, the inclusion of an equation of state $w\neq-1$ typically degrades
the parameter constraints as extra parameters generally increase parameter
uncertainties overall. (It is not surprising that an extended parameter space
is able to partially ease cosmological tensions.)
Though this gives a more general description of the dark sector
we note that the freedom enabled by a variable DE equation of state will to
some extent be captured by the model-independent form of $Q$ considered here.
We leave this more comprehensive analysis to future work when better data become
available.

As a caveat to using the CPZ correlated prior \eqref{eq:pi-q}
we are forced to assume a fiducial model of $q(a)$, which necessarily introduces
some bias into the reconstruction. Of course, some structure must be
introduced into the problem if we are to make progress. We have
obtained results against two fiducial models: one with a running average of $\bm{q}$
as in \cite{Crittenden:2011aa}, and a more conservative choice that favours
null interaction with $\qfid=0$. Reassuringly, features of the 
reconstruction where data are abundant ($z\lesssim0.4$) are largely the
same meaning that the choice of $\qfid$ is mostly irrelevant. By contrast,
going out to higher redshifts we find the reconstruction is prior
dominated, being highly dependent on the choice of $\qfid$.

The next generation of surveys will certainly allow a more elaborate,
fine-scale reconstruction of the interaction history than considered here.
A Fisher forecast shows that the constraints
from low-redshift data can be expected to improve by a factor of two, for
a minimal one parameter interaction model.
In summary, current data do not reveal any statistically
significant deviations from \LCDM. If however there is interaction to
be found at the level $q\gtrsim 0.04$, we anticipate that a one year sample of LSST
SN data combined with DESI BAO and RSD data will be capable of distinguishing
from \LCDM\ at a statistical significance of $\simeq 3\sigma$.

\section*{Acknowledgements}
We thank Jan Hamann and David Parkinson for comments and discussions during
the early stages of this work, and Florian List for his reading and comments
on an early manuscript.
LD is supported by the Australian government Research Training Program.
KB acknowledges the support of the Australian Research Council through
the Future Fellowship FT140101270. This work has made use of the 
publicly available codes
{\tt CAMB} \cite{Lewis:1999bs},
{\tt emcee} \cite{ForemanMackey:2012ig}, and 
{\tt getdist}.\footnote{\url{https://github.com/cmbant/getdist/}}
The authors acknowledge the use of Artemis at The University of Sydney
for providing HPC resources that have contributed to the research results
reported within this paper.

\appendix

\section{On using standard fitting formulae\label{FF}}
In general no analytical form exists for $z_*$ and
$\zdrag$ --- though widely used fitting formulae have been given \cite{Hu:1995en}.
We justify the use of these fitting formulae for our interacting model
by noting the following.
The sound horizon depends on physics before the time of recombination
physics ($z\sim 1000$) going back to the big bang.
While the standard formula given usually assume \LCDM\ the fact that
the interaction considered here is a late-time effect meaning that
the evolution tracks \LCDM\ up until $z\lesssim 2$. Thus, up to the
time of last scattering DM and DE evolve in their usual way.
In non-interacting models $z_*$ has a weak dependence on $\omb$ and $\omc$;
we find in our interacting model that it also has a weak dependence on $q$, 
as well.
Moreover, in standard recombination we have
\be
r_s(z)
\propto
\ln\left[\big(\sqrt{R(z)+R_\mrm{eq}}+\sqrt{R(z)+1}\big)/\big(1+\sqrt{R_\mrm{eq}}\big)\right],
	\label{eq:rs}
\ee
where $R(z)\equiv \dot\rho_b/\dot\rho_\gamma$ and
$R_\mrm{eq}= R(z_\mrm{eq})$, i.e.\ evaluated at matter-radiation equality
$z_\mrm{eq}$ (which is in general different from \LCDM).
The logarithm of \eqref{eq:rs} is order one and insensitive
to cosmological parameters, because the sound horizon is determined
largely by the physical densities $\omb$ and $\omc$.

\begin{figure}
	\centering
	\includegraphics[width=0.6\linewidth]{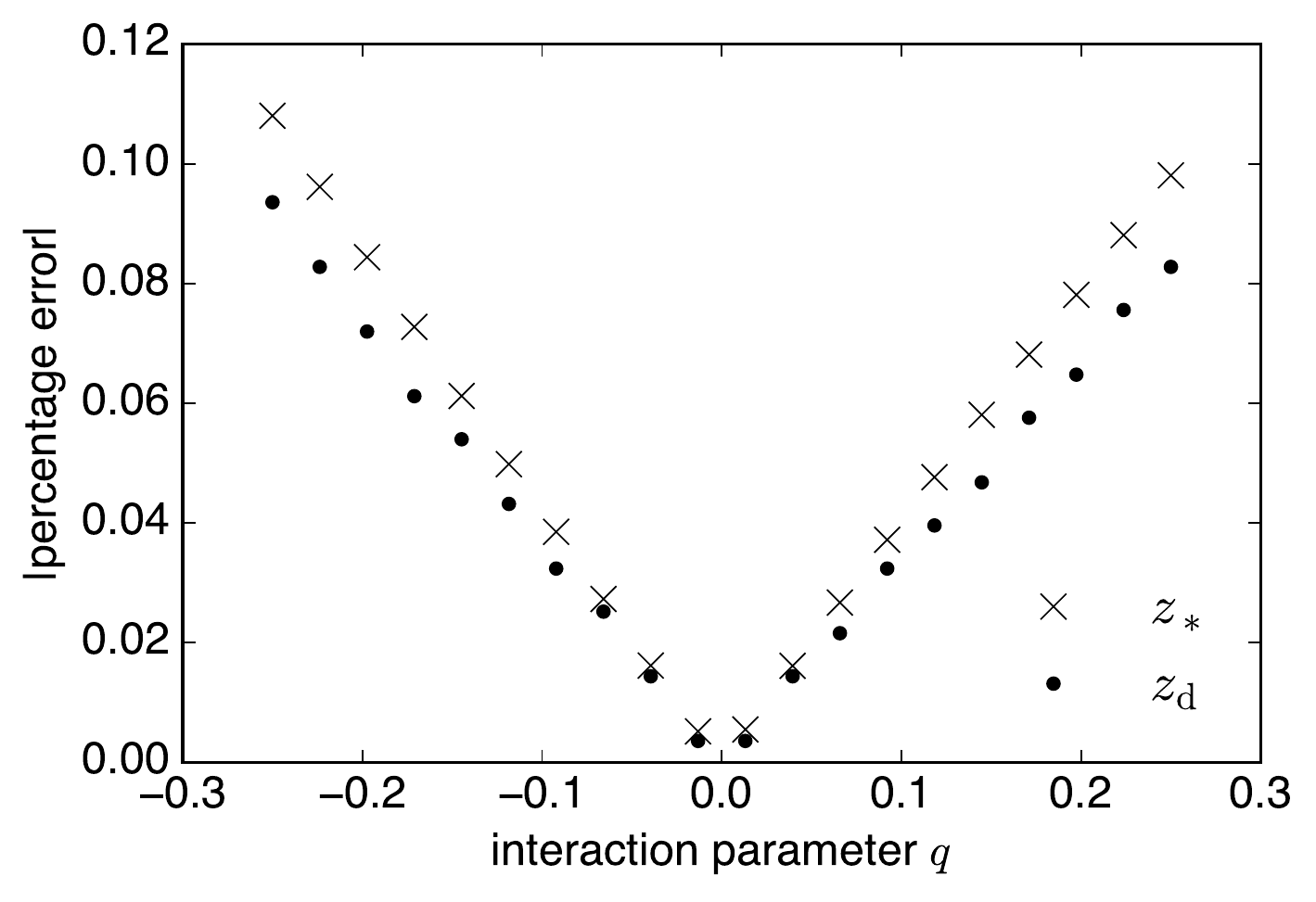}
\caption{The error $100\times |z-z_\mathrm{HS96}|/z_\mathrm{HS96}$
		for the redshifts $z_*$ and $\zdrag$ when using the Hu and Sugiyama
		fitting formulae \cite{Hu:1995en} as compared with the exact CAMB
		calculation \cite{Lewis:1999bs}.}
  \label{fig:zdec}
\end{figure}

In Figure \ref{fig:zdec} we have confirmed using CAMB \cite{Lewis:1999bs}
that the error for $\zdrag$ and $z_*$ is less than $0.1\%$ for a wide range
of $q$. Although this is for a single interaction parameter, we expect
this approximation to also hold for models with multiple $q$,
provided they are all of similar magnitude.

\section{Details of the SNe Ia likelihood\label{SN-BHM}}
The commonly used chi-squared method has been shown to be inadequate,
leading to a systematic shift in the cosmological parameters \cite{March:2011xa}.
This shift is due to the uncertainties of the colour parameter $c$ being
roughly of the same size
as its scatter \cite{gull:1989} and the issue originates from the implicit
flat priors on the stretch and colour parameters. The solution is to
adopt informative priors  and below we sketch the construction of the
likelihood using a recent Bayesian hierarchical approach.

The dependence of the observables on the latent variables can be 
made explicit by writing the likelihood as
\be\label{eq:LSN-marg}
\begin{split}
\mathcal{L}_\mrm{SN}({\theta};D)
&\equiv p(\hat{\bm{z}},\hat{\bm{m}}_B^*,\hat{\bm{x}}_1,\hat{\bm{c}} \,|\, {\theta}) \\
&=\int
p(\hat{\bm{m}}_B^*,\hat{\bm{x}}_1,\hat{\bm{c}} | \bm{m}_B^*,\bm{x}_1,\bm{c},{\theta}) \\
&\quad\times p(\bm{m}_B^*,\bm{x}_1,\bm{c} | \bm{M}, {\theta}) 
\times p(\bm{M} | {\theta})
\:d\bm{M} \, d\bm{m}_B^* \, d\bm{x}_1 \, d\bm{c}.
\end{split}
\ee
Note that we have ignored measurement errors in the redshift as it has
little effect on parameter estimation \cite{March:2011xa}.
(The redshift errors, however,
are still propagated to the apparent magnitude using \eqref{eq:distmod}.)
Thus we put $\hat{\bm{z}}=\bm{z}$ and have suppressed the
dependence in redshift.
Further conditioning on the latent stretch and colour
parameters we have
\be\nonumber
p(\bm{m}_B^*,\bm{x}_1,\bm{c} | \bm{M}, {\theta})
=p(\bm{m}_B^* | \bm{M}, \bm{x}_1,\bm{c}, {\theta})
	\, p(\bm{x}_1,\bm{c}|{\theta}).
\ee
Since $\bm{x}_1$ and $\bm{c}$ can be considered independent variables
the joint prior is separable, i.e.\
$p(\bm{x}_1,\bm{c}|{\theta})=p(\bm{x}_1|{\theta})p(\bm{c}|{\theta})$.
As $p(\bm{m}_B^* | \bm{M}, \bm{x}_1,\bm{c}, {\theta})$ 
expresses the deterministic relation \eqref{eq:tripp} we have
\be\nonumber
p(\bm{m}_B^* | \bm{M}, \bm{x}_1,\bm{c}, {\theta})
=\delta\big(\bm{m}_B^*-\bm{m}_B^*(\bm{M},\bm{x}_1,\bm{c};{\theta})\big),
\ee
where $\delta$ is the Dirac delta function and
$\bm{m}_B^*(\bm{M},\bm{x}_1,\bm{c};{\theta})
=\bm\mu(\bm{z};{\theta})+\bm{M}-\alpha\bm{x}_1+\beta\bm{c}$.
Marginalising over $\bm{m}_B^*$ \eqref{eq:LSN-marg} becomes
\be\nonumber
\begin{split}
\mathcal{L}_\mrm{SN}({\theta};D)
=&\int
p(\hat{\bm{m}}_B^*,\hat{\bm{x}}_1,\hat{\bm{c}} | 
	\bm{m}_B^*(\bm{M},\bm{x}_1,\bm{c};{\theta}),\bm{x}_1,\bm{c},{\theta}) \\
&\times p(\bm{M} | {\theta})\, p(\bm{x}_1 | {\theta})\, p(\bm{c} | {\theta})
\:d\bm{M} \, d\bm{x}_1 \, d\bm{c}.
\end{split}
\ee

The SALT2 outputs for the $i^\mrm{th}$ SN Ia $\{\hat{m}_{B,i}^*$, $\hat{x}_{1,i}$,
$\hat{c}_i\}$ are correlated variables and so we construct
$p(\hat{\bm{m}}_B^*,\hat{\bm{x}}_1,\hat{\bm{c}} | \bm{m}_B^*,\bm{x}_1,\bm{c},{\theta})$
to be a $3N$-dimensional multivariate Gaussian with mean given by the
corresponding latent variables ($\bm{m}_B^*$, $\bm{x}_1$ and $\bm{c}$)
with a $3N\times 3N$ covariance block diagonal matrix $\Sig_\mrm{stat}$.
Each SN Ia, having latent variables $M_i$, $x_{1,i}$ and $c_i$, are 
plausibly assumed to be drawn from independent and identical Gaussian
distributions, i.e.\ $M_i\sim\calN(M_0,R_M^2)$,
$x_{1,i}\sim\calN(x_*,R_x^2)$ and $c_i\sim\calN(c_*,R_c^2)$. With 
these considerations the integral \eqref{eq:LSN-marg} reduces to
a convolution of Gaussians, which can be analytically resolved
to give the marginalised negative log-likelihood
\be\label{eq:LSN}
\begin{split}
-\ln\calL_\mrm{SN}({\theta};D)
=\frac{1}{2}
	(\bm{\widehat{Y}}-\mat{B}\bm{X}_0)^\T
	(\mat{B}\Sig_\mrm{sys}\mat{B}^\T + \Sig_\mrm{stat})^{-1}
	(\bm{\widehat{Y}}-\mat{B}\bm{X}_0)& \\
+\frac{1}{2}\ln\det(\mat{B}\Sig_\mrm{sys}\mat{B}^\T + \Sig_\mrm{stat})
+\mrm{const}, &
\end{split}
\ee
where
\begin{align*}
\bm{X}_0 &= (M_0,x_*,c_*,M_0,x_*,c_*,\ldots,M_0,x_*,c_*)^\T, \\
\widehat{\bm{Y}} &= (\hat{m}_{B,1}-\mu_1,\hat{x}_{1,1},\hat{c}_1,
				\hat{m}_{B,2}-\mu_2,\hat{x}_{1,2},\hat{c}_2,
				\ldots,
				\hat{m}_{B,N}-\mu_N,\hat{x}_{1,N},\hat{c}_N)^\T,
\end{align*}
are vectors of length $3N$, $\Sig_\mrm{stat}$ is the covariance matrix
of statistical uncertainties from the light-curve fit, and $\Sig_\mrm{sys}$ is
the covariance matrix of systematic uncertainties, including from
calibration, the light-curve model, dust extinction, and bias
uncertainty \cite{Betoule:2014frx}. Finally,
$\mat{B}=\mathrm{diag}(\mat{J},\mat{J},\ldots,\mat{J})$
is a $3N\times 3N$ block diagonal matrix where each block is identical
with
\be
\mat{J}=
\begin{pmatrix}
1 & -\alpha & \beta \\
0 & 1 & 0 \\
0 & 0 & 1
\end{pmatrix}.
\ee
Note $\Sig_\mrm{stat}$ has parameter dependence so the normalisation term of
\eqref{eq:LSN} cannot be neglected in MCMC parameter estimation. Moreover
despite the Gaussian form of \eqref{eq:LSN} the data
$\{\hat{\bm{m}}_B^*,\hat{\bm{x}}_1,\hat{\bm{c}}\}$ are not 
Gaussian distributed.
For a more detailed derivation, including analytic marginalisation of
hyperparameters, we refer the reader to \cite{March:2011xa}.


\end{document}